# The Arrival of Fast Internet and Employment in Africa
## Comment

David Roodman*

Open Philanthropy

September 2025

**Abstract**

Hjort and Poulsen (2019) frames the staggered arrival of submarine Internet cables on the shores of Africa circa 2010 as a difference-in-differences natural experiment. The paper finds positive impacts of broadband on individual- and firm-level employment—with a bias toward skilled employment—and on nighttime light emissions. These results largely are not robust to alternative geocoding of survey locations, to correcting for a satellite changeover at end-2009, and to revisiting a definition of the treated zone that has no clear technological basis, is narrower than the spatial resolution of nearly all the data sources, and is empirically suboptimal as a representation of the geography of broadband.

## Introduction

Hjort and Poulsen (2019)—HP19 for short—is the first work to estimate the distributional impacts in poor countries of a technology thought to have increased inequality in rich ones (HP19, p. 1036). HP19 obtains its evidence by framing a difference-in-differences (DID) design around the landing of ten undersea data cables on the shores of Africa in 2009–12. Before then, the continent's Internet bandwidth was constrained by older cables and satellite links. The removal of the bottlenecks surely delivered higher data speeds to households and firms served by national fiber optic networks. The cable connection events thus generated treatment variation over both space and time. Given the uncertainties in large engineering projects, the timing component in particular was plausibly, locally exogenous—exogenous, that is, at the scale of a year.

* david.roodman@openphilanthropy.org. Thanks to Jonas Hjort for assistance with data and code and feedback on an earlier draft, to Otis Reid for guidance and ideas, and Steve Song for advice. The packages esttab (Jann 2007), blindschemes (Bischof 2017), grc1leg2 (Over 2022), reghdfe (Correia 2014), and reghdfejl (Roodman 2024) were used to estimate, tabulate, and plot. The nighttime light data used here is produced by Earth Observation Group, Payne Institute for Public Policy, Colorado School of Mines, using DMSP data collected by US Air Force Weather Agency. 3G coverage data © Collins Bartholomew and GSMA, used with permission. Replication data and code at https://doi.org/10.7910/DVN/V7ASIP.

HP19 traces the impacts of this treatment in many data sets, most of which are gathered here too: city-level bandwidth measurements from the content delivery company Akamai; multi-country household surveys from Afrobarometer and the Demographic and Health Surveys (DHS) program; South Africa's Quarterly Labour Force Surveys (QLFS); Ethiopia's firm-level Large and Medium Scale Manufacturing Industries Survey (LMMIS); the World Bank's multi-country enterprise surveys (WBES); and nighttime light observations from the Defense Meteorological Satellite Program's Operational Linescan System (DMSP).

HP19 finds positive impacts from broadband access in all the data sets. The collective coherence of the results adds to their individual credibility. Local broadband service increases the probability of an individual being employed by 4.6 percentage points in the DHS follow-ups, 7.7 points in the Afrobarometer surveys, and 2.2 in the South Africa QLFS. The effect is positively skill-biased, meaning that it is largely confined to employment in skilled occupations. Yet, counterintuitively, the effect size is comparable for less- and more-schooled individuals. Separately, broadband access is associated with 2–3% higher nighttime light emissions, which is taken to indicate increased economic activity (Chen and Nordhaus 2011; Henderson, Storeygard, and Weil 2012). HP19 also finds positive impacts on firm entry, employment at existing firms, and firm exports.

Much of the work in HP19 relates to the geographic dimension of the data:

constructing maps of backbone networks circa 2010, estimating coordinates for localities identified in survey data only by place name, and classifying locales as "treated" or "untreated" according to proximity to digital infrastructure.

This comment mainly revisits that geographic work. One step I take is to re-estimate the geographic coordinates of observations in primary sources whose locations are given by name. I recode the Akamai and Ethiopia LMMIS data using public sources such as Google Maps. For Afrobarometer enumeration areas, I turn to the third-party geocoding by AidData and Afrobarometer (BenYishay et al. 2017).

I also revisit the definitions of the treatment and control groups. HP19's base specifications define the treated zone as the area within 0.005° of a fiber optic line, which is about 560 meters. (HP19 does vary this parameter in a robustness test.) The control zone is the area within 0.005–0.1°, or up to about 11 kilometers. The 0.005° threshold invites critique from a few directions. It is not rooted in technical characteristics of communications equipment. Wired and wireless means deliver broadband well more than 560 meters from the backbone. And the threshold produces an asymmetry, a control zone 19 times wider than the treatment zone. The precision in the definition is in a sense illusory since almost none of the location measurements in the HP19 data, including the paths of the fiber lines are reliably estimated to the sub-kilometer scale. As mentioned, many of the primary sources identify locations only by naming localities or districts, which can be well more than 560 meters wide. The DHS data includes numerical latitude and longitude coordinates, but injects kilometer-scale noise for privacy. Even the satellite data may contain comparable measurement error (Elvidge et al. 2004), and at any rate HP19 downsamples the light data to 0.1° squares. In effect, by preferring a narrow treatment band, HP19 attains its results after reducing the power to detect them. To the extent that the research and public process favor statistically significant results—however unconsciously—lower power is associated with higher propensity to generate significant but non-representative results (Ioannidis, Stanley, and Doucouliagos 2016). And while the measurement error attenuates exogenous identifying variation, it does not attenuate endogenous influences in tandem, so they gain relative importance.

Acting on suggestions from the experts who supplied the underlying data on the geography of the African backbone network, several alternative representations are proposed here, and tested for fidelity: defining the treatment zone as 0.05° from fiber lines instead of 0.005°, while retaining the 0.1° radius for the control zone bound; doubling both those numbers, to 0.1° and 0.2°; and retaining the latter bounds while taking distance from the nearest network *node* rather than the nearest line.[1] The motivation for taking distance from nodes is that HP19's line-based representation causes localities in "digital flyover country" to be coded as treated because fiber lines run nearby, regardless of whether they provide broadband locally.

Empirically, the alternatives more accurately represent the geography of broadband. For example, using an early (2014) mobile phone 3G coverage map, the $t$ statistic for a difference in broadband access across the HP19-preferred treatment and controls zones is 1.83 in the DHS sample. For the HP19-geocoded Afrobarometer locations, $t = -4.00$: the rate of 3G coverage was *lower* in the HP19 treated sample. Switching to the 0.1° and 0.2° bounds for treatment and control zones (and to the Afrobarometer-endorsed geocoding) produces corresponding $t$ statistics of 10.82 for DHS and 10.38 for Afrobarometer. That indicates that the wider zones represent the geography of broadband with more fidelity, and so should undergird higher-powered impact estimation.

[1] In conversations, Paul Hamilton of Africa Bandwidth Maps, and Steve Song, creator of the AfTerFibre fiber network map, preferred defining treatment on the basis of distance from network nodes rather than fiber lines, and bounding the treatment zone at 0.05° or 0.1° rather than 0.005°. In light of the diversity of local circumstances and incompleteness of the available data, neither endorsed these proposals with confidence.

Lastly, this comment introduces three changes in the nighttime light analysis: taking the data at its original 1/120° resolution instead of a downsampling to 0.1°; adding a control to prevent confounding from a satellite changeover between 2009 and 2010; and switching to an "intercalibrated" data set that makes the data more comparable over time.

Most HP19 findings shrink in magnitude and statistical significance under these revisions. A meta-analytic average of the three estimates of the impact on individual-level employment (using DHS, Afrobarometer, and South Africa QLFS data) drops from 3.6 points (standard error 1.2 points) for the original results to 2.0 (0.7) after making a few data and specification corrections and revising the Afrobarometer geocoding. When using higher-fidelity treatment and control zone classifications, it falls further, to 1.2 (0.9) or 0.3 (0.6) points. The main source of positivity in these results is the DHS data. Yet that family of surveys is the least frequent of the ones used here, with an average gap between pre- and post-event surveys of almost 9 years. Its identifying temporal variation is thus dominated by long-term trends that are, while more important than short-term changes, also more presumptively endogenous. Meanwhile, the findings of skill bias disappear.

As to firm-level impacts, in the regressions on World Bank Enterprise Survey data, identification turns out to flow solely from the Nigeria data, which are geographically labeled only by state and thus contain almost no information about proximity to fiber networks. The Ethiopia firm-level results are robust to re-geocoding, but not to widening the treatment and control zones.

HP19 runs two regressions on nighttime light data. The result from the first be explained by confounding from a satellite changeover between 2009 and 2010, which altered the measurement regime. The second regression adds fixed effects that remove that confounding, but in such a flexible way that identification no longer flows from the paper's quasi-experiment.

Section 1 reviews the HP19 specifications. Section 2 discussed the HP19

definitions of treatment and control zones and details the re-geocoding of three data sets. Sections 3 and 4 present results for non-satellite and satellite data. Section 5 concludes.

## 1 The HP19 specifications

HP19 estimates impacts on a variety of outcomes, through specifications of a shared form, linear two-way fixed effects (TWFE). The predictor of interest is the product of temporal and cross-sectional indicators, namely, a dummy for whether an observation occurred after a country received undersea cable connection between 2009 and 2012, and a dummy for whether the observation occurred proximate to a fiber line thought active circa 2010. The idea is that, within each country, broadband access was distinctly greater after a connection event, and more so for locales close to the national fiber backbone.

The data sets to which TWFE is applied differ in their cadence. The multi-country DHS, Afrobarometer, and WBES surveys are infrequent in any given country. So from these sources HP19 uses one "before" and one "after" round for each country.[2] The average of the country-level before-after gaps is almost 9 years for the DHS, 4 for Afrobarometer, and immaterial for the WBES, as will be explained. The Ethiopia firm survey and night light data are annual, and the Akamai Internet speed and South Africa labor force data quarterly. From these higher-frequency sources, multiple rounds are taken.

The HP19 specifications control for several sets of fixed effects. Since the predictor of interest is the product of geographic and temporal indicators, the TWFE design requires the control set to cover those two factors separately, with sets of fixed effects. When working with the pooled data from multi-country surveys, HP19 goes farther, interacting the temporal indicators (year, quarter, or post-treatment dummies) with country dummies. The regressions thus pool country-specific TWFE quasi-experiments while subjecting them to the restriction of equal impacts. To reduce spatial autocorrelation, HP19 typically adds more fixed effects, for enumeration areas (South Africa QLFS and Ethiopia LMMIS), cities (Akamai and WBES), or interactions between the treatment dummy and $0.1°\times0.1°$ grid cell dummies (DHS and Afrobarometer).

[2] There is one exception: the WBES data includes three rounds from Mali.

When working with multi-country data sets (Akamai, DHS, Afrobarometer, DMSP-OLS), HP19 sometimes adds another set of fixed effects to test robustness, in a way that moves from a double- to a triple-differenced design. The added effects are interactions between the treatment zone dummy (called "connected" in HP19) and time dummies. The idea is to "control for a nonlinear trend in employment [or other outcome] that is specific to the connected locations." Controlling for every combination of treatment and time period is analogous in classical 2×2 DID to including a dummy for each of the four cells. Put otherwise, the HP19 treatment—being close to fiber after it is hooked up to broadband—is one "trend in average incomes that is specific to the connected locations" that is controlled away by the extra fixed effects.

Nevertheless, the coefficient on treatment may still identified in multi-country regressions. What is estimated is not the effect of gaining broadband within the treated zone, relative to not gaining it in the untreated zone (a double difference), but the difference between that double difference for a country that got broadband earlier and the double difference for a country that got broadband later (which is a triple difference). It is not obvious that the tripled-differenced specification can be viewed as quasi-experimental, for it leans heavily for identification on when various countries received the treatment, and that was not necessarily arbitrary. It is unsurprising that the richest country in the study, South Africa, was in the first batch of countries to start the treatment, in 2009.

In fact, HP19 reports results from triple-differenced specification on two data sets for which it is not actually identified, DHS and Afrobarometer (Table 2, columns 5 and 7; Table 4, columns 2 and 4). Here, the treatment indicator is collinear with the fixed effects, but the estimation software drops one of the fixed effect dummies rather than the treatment indicator.[3] For the HP19 triple-difference estimator to be identified, there must be at least one year in which a) at least two countries are surveyed and b) at least one is still pre-treatment and at least one is post-treatment. Otherwise, there is no effective cross-country difference in the time profiles of treatment. The DHS data has observations from 0 countries in 2009, 1 in 2010, 0 in 2011, and in 2012. Only in 2013 are multiple countries surveyed, and by then all are post-treatment. And before 2009, all countries are pre-treatment. The story is similar in the Afrobarometer data.

HP19 is correct that the triple-differenced specification is "unusually demanding of the data." In the DHS and Afrobarometer data, it is not identified. In the Akamai and DMSP (satellite) data, it is identified, but its causal interpretation is less clear.

## 2 Revisiting the geographic information and definitions

### 2.1 Reconstructing the distance variable

I obtain DHS, Afrobarometer, WBES, DMSP, and South Africa QLFS data sets from their primary providers. I take Akamai and Ethiopia LMMIS data from the HP19 archive. I revisit the geographic processing of all of the data sets. Several revisions result:

1. Since distances in HP19 were computed with the Pythagorean formula, treating latitudes and longitudes as coordinates in the plane, I recalculate them in a standard geographic coordinate system (World Geodetic System 1984).[4] This most affects observations from South Africa since it is farthest from the equator.
2. The AidData geocodings of the Afrobarometer household survey locations (BenYishay et al. 2017) are used in place of the HP19 geocodings.

[3] This behavior is avoided if the treatment zone–year dummies are moved to the absorb() option of the areg, reghdfe, or reghdfejl command.

[4] Earlier versions of this comment committed the same error.

AidData, based at the College of William and Mary, has long experience in geocoding data from developing countries. The coding project was carried out by a "team of trained geocoders" according to a "double-blinded" process in which two people independently coded each location, disagreements went to an arbitration round, and quality was checked in other ways. Here, only geocodes scored by BenYishay et al. as fully precise (precision level 1) are retained.

3. The 77 localities sampled by the Ethiopia LMMIS are manually re-geocoded by drawing on publicly available GIS data on towns and woredas (third-level administrative units).[5] The LMMIS data set in the HP19 public archive names all three administrative levels for each observation.
4. Most localities contributing identifying information in the Akamai Internet speed data set are similarly recoded. These localities are those that satisfy HP19's criteria of providing more than 10 speed measurements in at least one quarter; and being in countries with data adequate for TWFE, i.e., ones with at least one treatment city and at least one control city with pre- and post-treatment data. OpenStreetMap's Nominatim engine is the primary source of coordinates, and Google Maps a secondary one.[6] 168 of the 186 localities are coded.
5. The DMSP light data is taken at its native resolution of 1/120° rather than downsampled to 1/10°. An intertemporally calibrated version of the data is also analyzed, as explained in section 4.

[5] Town coordinates are taken from the Google Maps API. For Woredas, coordinates of prominent towns on major roads are used, since they are better guesses for the locations of large and medium-sized firms.

[6] The Nominatim engine often returns multiple results. In most cases, I use the result having *class* = "place" with highest *importance*. In the case of ties—typically Nigerian villages with the same name—I do not geocode unless Google Maps favors one locality.

## 2.2 Defining treatment and control groups

As in the human body, a fiber optic backbone supports a filigreed network. At junctions called nodes, feeders branch out from the backbone, in a variety of technological forms. These technologies include asymmetric digital subscriber line (ADSL), which runs over copper phone lines; Data Over Cable Service Interface Specification (DOCSIS), which does the same on cable television wires; local fiber lines ("fiber to the home"); fixed wireless, which is essentially high-powered Wi-Fi; WIMAX, the mesh-network equivalent; microwave, for long-distance transmission; and cellular towers, which are considered to supply broadband in generation 2.5G or higher and which are themselves linked to the backbone through wired or wireless means. Most broadband in Africa is delivered wirelessly. In 2010, Africa had 1 million wired broadband subscriptions and 14 million wireless ones. By 2021, the totals had risen to 7 million and 448 million.[7]

**The HP19 treatment and control zone definitions**

To estimate the impacts of broadband in Africa under a dichotomous definition of connectivity as in HP19, one needs to classify observations according to whether they have access to high bandwidth in the post-treatment period. It is not a straightforward question how best to do so, given the complex and rapidly changing anatomy of the continent's networks, and the limited data thereon circa 2010. This section describes HP19's approach, comments on it, and proposes alternatives. It will be argued that the alternatives more accurately represent the geography of broadband in the study setting, and so endow impact regressions with greater power. And switching to evidence-backed alternatives tests robustness in a minimally arbitrary way.

HP19's classification proceeds in three steps. First, the decision is made to proxy broadband availability by *distance from fiber infrastructure*. Finer-grained information on mobile phone coverage or distance from municipal fixed-line networks would be preferable, but is scarce for the study setting.[8] (For many African countries, 3G coverage data only begins in 2014.) Second, fiber optic *lines* are chosen as the distance referent. Third, *bounds* (radii) are set to demarcate the treatment and control groups. The treatment zone is within 0.005 angular degrees from the lines and the control zone spans from there to 0.1°. At the Earth's surface, each angular degree extends about 111.32km, so 0.005° and 0.1° are approximately 0.56km and 11km.

[7] See International Communications Union data set, itu.int/en/ITU-D/Statistics/Documents/facts/ITU_regional_global_Key_ICT_indicator_aggregates_rev1_Jan_2022.xlsx, "By BDT region" tab.

One downside of HP19's second step, taking fiber lines as the referent, is that some people live and work in digital flyover country—near the lines, but not connected to them. Metaphorically speaking, they can watch the bits fly by, but not send or receive them. *Access* requires connection to a network node via wired or wireless feeders. Since Africans usually access broadband through cell towers, many of which are erected near fiber nodes, distance to the nearest node may correlate better with bandwidth availability. For this reason, I introduce a geocoded fiber node data set from Africa Bandwidth Maps (ABM).

Whether a given node-based representation coheres better with the geography of broadband access than a given line-based representation is an empirical question, which will be explored just below. Many cell towers, which are the true point sources for most broadband in Africa, are not right at fiber nodes. Meanwhile, if households and firms congregate along the major roads that fiber lines also follow, and their purchasing power attracts cell towers, a line-based

[8] Related studies, in later years or other countries, use cellular coverage data. Collins Bartholomew coverage maps feature in Manacorda and Tesei (2020), Masaki, Ochoa, and Rodríguez-Castelán (2020), Gonzalez (2021), Guriev, Melnikov, and Zhuravskaya (2021), Chiplunkar and Goldberg (2022), Goldbeck and Lindlacher (2022), and Mensah, Tafere, and Abay (2022). Klonner and Nolen (2010) uses a South Africa–specific source. Bahia et al. (2020) models coverage using information on cell towers.

representation of broadband access may in fact be reasonably accurate.

With more confidence, this reanalysis proposes an alternative to the choices made in the third step of the HP19 classification, the delimitation of the treatment and control zones by 0.005° and 0.1° radii. The initial version of HP19 places no outer bound on the control zone and sets the treatment boundary at the *median* distance from the network, which is about 19.7km for the Akamai data set, 37.5km for DHS, 2.7km for Afrobarometer, 1.04km for the Ethiopia LMMIS, 3.4km for the South Africa QLFS, and 3.8km for WBES (Hjort and Poulsen 2016, note 36). The final version (HP19, note 34) justifies the switch to the much narrower 0.005° treatment bound as follows: "Technological considerations indicate that 500 meters is a reasonable proxy for potential fast Internet reach beyond the backbone cables for copper-cable last mile technologies."

The 0.005° marker appears hard to justify. HP19 does not specify the technological considerations it mentions. Perhaps the reasoning behind the emphasis on copper in HP19's explanation is that wired connections, where available, exercise disproportionate economic influence. But, as already noted, wireless access dominates by far in Africa, even back in 2010. It is central to three of HP19's five anecdotal examples of impact (HP19 §I.C): rising e-commerce, domestic manufacture of low-end Internet-capable devices, and improved coordination among farmers and other supply chain actors. To the extent that the reach of copper networks matters, copper can in fact reach much farther than 500 meters, as it does in cities with wired phone networks. The ADSL2 standard, promulgated in 2002, can deliver broadband up to 5–6km (Kagklis et al. 2005). And even if it could not, when the outcome is individual-level employment and the place of observation is the home, the economic influence of broadband would spread farther: broadband would reach firms within 500 meters, which would hire people living more distally from the lines.

A final concern is that the use of a 0.005° (~0.56km) bound implies an unrealistic expectation of precision in the data. Each download speed, household, or firm observation in HP19 is geocoded by an enumeration area or some larger locality such as a city, to which a representative coordinate pair is assigned. Most of the localities themselves extend more than 0.56km in one direction or another, the main exception being certain urban enumeration areas in the South Africa QLFS. Since all observations in each enumeration area or city are assigned the same coordinate pair, all are assigned the same treatment status, even when some might be considered treated and some not, if their true distances from fiber infrastructure were known. Distinctively, the DHS provides actual coordinates for the centroids of enumeration areas, not just names. But it randomly perturbs the coordinates for privacy—by 0–2 km for urban observations, 0–5km for 99% of rural ones, and 0–10km for the other 1% (Perez-Heydrich et al. 2013).[9] Even the satellite data may contain kilometer-scale error: Elvidge et al. (2004, p. 290) finds positional errors of 1.55–2.36km in DMSP readings, at least when taken from a single orbit, rather than averaging together many as is done for the annualized DMSP data used in HP19. At any rate, HP19 downsamples the light data to 0.1° (~11km) resolution, from its native resolution of 1/120°. Finally, the fiber paths are digitized with error too. Since more than 90% of fiber lines in Africa run along roads,[10] the routes in the AfTerFibre database that is the basis for the HP19 network map are usually derived from vector representations of major roads, which do not capture all bends and curves.[11] Figure A–1 in the appendix illustrates with the case of a line running through Cotonou, Benin. As digitized, the line deviates up to a kilometer from the curved national highway that it probably runs along.

[9] The DHS perturbs reported survey locations by drawing from uniform distributions for the angle and distance of perturbation. Under this process, if the maximum perturbation distance is $R$ then the average absolute perturbation along a given axis is $R/\pi$. If a survey location were exactly on a fiber line, if that line's path were locally straight, and if the recorded path did not also contain measurement error, then the expected estimated distance from point to line would be $R/\pi$.

[10] Conversation with Paul Hamilton, Africa Bandwidth Maps, August 30, 2022.

[11] AfTerFibre database, AfTerFibre.carto.com/tables/af_fibrephase/public, accessed June 11, 2022.

Giving the treatment zone a width at or below the resolution of most of the geographic data injects substantial error into the HP19 treatment assignments. One might reasonably share HP19's expectation (note 34) that this error makes the estimation strategy conservative. Measurement error in a right-side variable attenuates its coefficient estimate. However, that result assumes that treatment and the measurement error are exogenous. In observational data on complex social systems, in which causal arrows run in all directions, such assumptions are not strictly true, even after double-differencing and adding controls. In general, the pragmatic hope is that in a well-framed design, biases from violations of identifying assumptions are second-order. However, as the treatment zone narrows beyond the resolution of the data, erstwhile second-order biases can gain relative importance. Appendix B demonstrates this claim rigorously. Moreover, reducing precision can combine in a distortionary way with conscious or unconscious biases within the research and publication process toward statistically significant results, raising the risk that published results are unrepresentative of the sampling distributions they are described as coming from, and thus potentially fragile (Denton 1995; Gelman and Loken 2013; Ioannidis, Stanley, and Doucouliagos 2016).

**Alternative treatment and control zone definitions**

To ground an exploration of alternative bounds, I first illustrate the geography of broadband in Africa with two maps, one of a city, one of a country. Figure 1 depicts the Nairobi region. A burst of technology start-ups there constitutes one success story in HP19, in which wired broadband was plausibly crucial.

The top layer of the map, in orange squares, shows where the Internet speed testing company Ookla measured wired connections exceeding 256 kilobits per second (Kbps), the official definition of broadband in the study period (ITU 2010, p. 13). The data source starts in the first quarter of 2019, making that the period closest to the HP19 study setting, within the limits of data availability.[12] A layer down, in green, are locations of Ookla *wireless* broadband measurements. The lowest layer shows 3G coverage from a different data set, which is maintained by Collins Bartholomew on behalf of the international industry association GSMA. Mobile phone network operators supply data to it voluntarily—or don't. Substantial information first appears in HP19 study countries, aside from South Africa, in 2014, so that year's values are mapped. 2014 is also the follow-up year for Kenya in HP19's DHS sample. Collins Bartholomew asks network operators to distinguish "strong" coverage (signal strength above –92 decibels) from "variable" coverage, which may be unreliable indoors. Those coverage levels are shown in dark and light purple. The distribution of strong 3G coverage is taken to proxy availability of wireless broadband, as 2G is much slower and 4G was hardly available circa 2010 outside of urban South Africa.

The two data sources on the reach of broadband may be biased in opposite directions. The Ookla data set is a convenience sample, and so may understate access to broadband. Mobile network operators have an incentive to exaggerate the reach of their towers.

Finally, the map depicts the routes of fiber optic lines as recorded in the public HP19 data archive. The thin yellow ribbons with dashed borders show the HP19 0.005°-radius treatment zone while the wider yellow ribbons show the 0.1°-radius control zone.

The map of the Nairobi region supports certain conclusions. First, technological limits on wired connections did not confine broadband to within 0.005° of fiber lines. Indeed, HP19's control zone bound of 0.1° embraces much territory apparently reached by broadband in 2014. Second, the area of intensive broadband availability centers not so much on fiber lines as on a major junction in the network. To more accurately represent the distribution of broadband access with dichotomized treatment and control zones, it would seem reasonable to widen the bounds from 0.005° and 0.1° to 0.1° and 0.2°, and to switch their referent from network lines to network nodes. These wider bounds would have the added virtues of not implying an unrealistic demand for precision in the available data, and being minimally arbitrary, round numbers that give the same width to the treatment and control zones.

[12] Speedtest by Ookla Global Fixed and Mobile Network Performance Maps, registry.opendata.aws/speedtest-global-performance, accessed May 4, 2023.

The second map, of Ghana, provides a larger-scale check on the impressions from the first. It also adds blue circles to represent the 0.1°/0.2°-from-nodes classification just proposed, using the list of nodes from Africa Bandwidth Maps. (See Figure 2.) Again, HP19's treatment zone—depicted as the thin yellow spines within the wider yellow bands—appears to exclude much treated area. Yet the node-based definition also commits many type I and type II errors. If anything, it seems, a point-source-based representation would be more accurate *away* from fiber lines (and fiber nodes), where service is provided by isolated cell towers uplinked by microwave or satellite—assuming the data on point sources included such locations. Along the southern coast of Ghana, the range of 3G coverage parallels the fiber lines and is not confined to areas near nodes recorded in the data used here.

To check these casual impressions, I compute, for each of several possible definitions of the treatment and control zones, and each broadband proxy (Ookla measurements and "strong 3G" coverage from the GSMA), the $t$ statistic for the hypothesis that there is no treatment-control difference in the rate of broadband access. I compute the statistic separately for the geocoded locations in all four data sets in which employment effects will be estimated. Samples are subject to HP19 restrictions, such as age being below 65. The "strong 3G" calculations include only countries with substantial 3G data as of 2014, which are Ghana, Kenya, Madagascar, South Africa, and Tanzania. The $t$ statistics are computed for five classifications: the original HP19 data and definitions (0.005/0.1° from lines); a revised version that introduces certain geocoding and sample changes while retaining the HP19 radii (see section 3.3); and three novel alternatives, namely, 0.05°/0.1° from lines, 0.1°/0.2° from lines, and 0.1°/0.2° from nodes.[13] The last only factors in nodes active in 2010.[14]

The numerical results, all in Table 1, show that the alternatives improve on the HP19 classification. In the HP19 Afrobarometer sample, under the HP19 geocodings, the $t$ statistics are negative, meaning that the control zone had more broadband (last column of Table 1). Re-geocoding Afrobarometer place names improves matters, making the $t$ statistics positive (second row of the table). The 0.05°/0.1°-from-lines definition performs better in all data sets except the Ethiopia LMMIS. Doubling to the 0.1°/0.2°-from-lines classification produces the largest $t$ statistics in all data sets save the Ethiopia one (averaging $t$ statistics across the two broadband proxies). The 0.1°/0.2°-from-nodes classification performs not quite as well. In the rest of this paper, I will for concision omit impact estimates using 0.05°/0.1°-from-lines. I will focus on results from the two final classifications, 0.1°/0.2° from lines and nodes.

Two objections might be raised against this focus, though neither with much force. First, while a treatment classification that represents actual broadband access with more fidelity should increase power, the added power might come at a price in bias, as is often the case in econometrics. But there is little reason to expect that comparisons between the 0–0.1° and 0.1–0.2° zones should contain more endogeneity bias than ones between the 0–0.005° and 0.005–0.1° zones. Similarly, it might be objected that while the new classifications have higher fidelity in the cross-section, the same does not necessarily hold in the TWFE design. Perhaps the bandwidth *added* circa 2010 still accrued mainly within 0.005° of the fiber lines. But the coverage data benchmarked against here pertain to years just after treatment arrived, and the maps make it appear unlikely that technological factors limited the reach of additional broadband to 0.005° from fiber lines.

---

[13] The $t$ statistics are clustered in the same way as in corresponding regressions in section 3.4.

[14] HP19 models the fiber network as of 2009, which argues for using the node list of that year as well. However, 2010 is the first year in which all study countries—notably the DRC—have nodes in the ABM data.

**Figure 1. 3G coverage and points of high download speed around Nairobi, with line-based representations of broadband access**

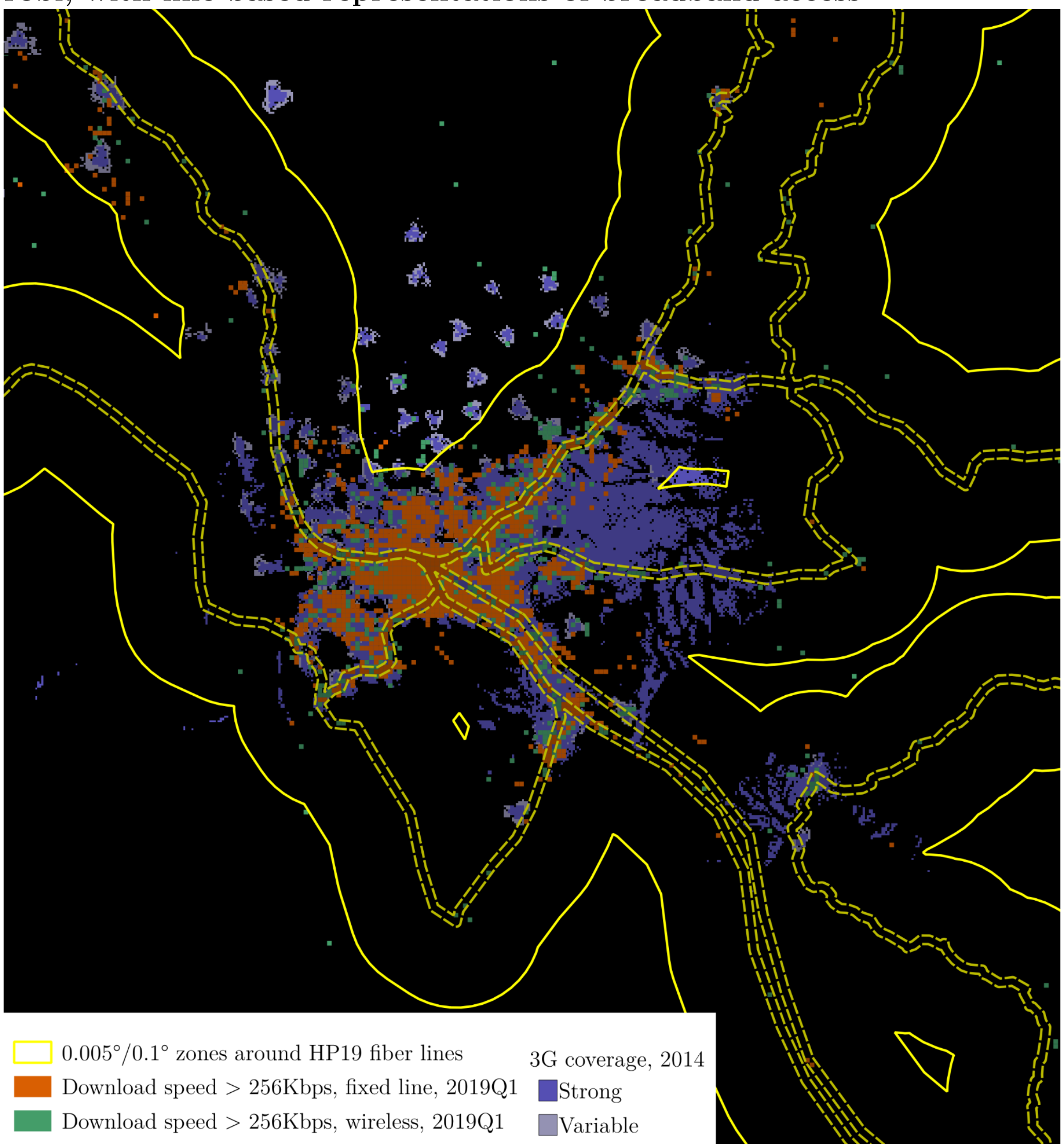

Notes: The base map shows, in shades of purple, 3G coverage in 2014. Orange (green) dots show locations with fixed (wireless) download speeds above 256 kilobits in the first quarter of 2019, according to Ookla. Dashed yellow lines depict the HP19 treated zone and solid yellow lines depict the control zone.

**Figure 2. 3G coverage and points of high download speed in Ghana, with line- and node-based representations of broadband access**

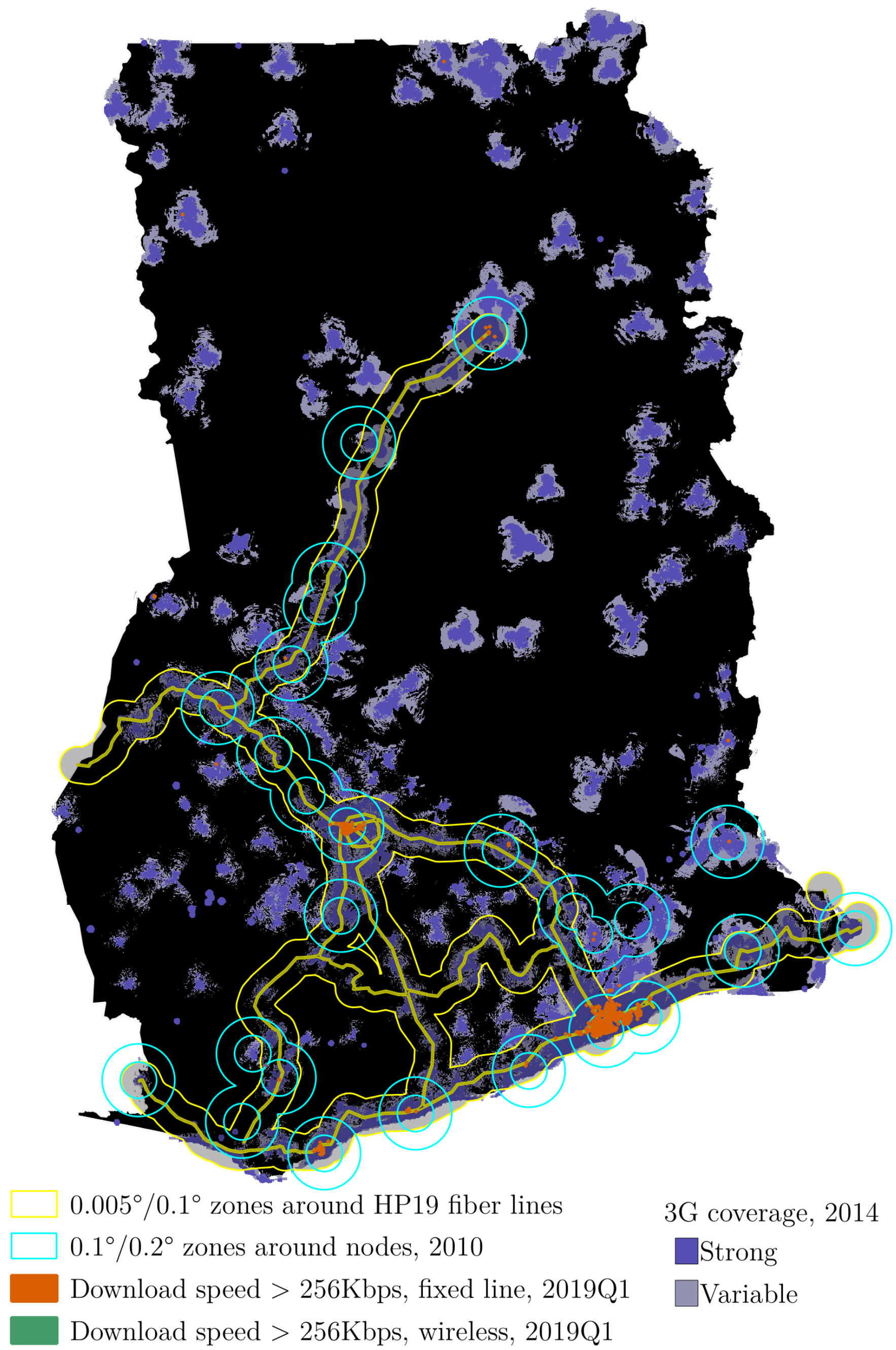

Notes: Symbology is the same as in Figure 1, with the addition of blue circles for node-based representation of broadband access. Node locations, for 2010, are from Africa Bandwidth Maps.

**Table 1. Coverage rates of broadband under various proxies for broadband and definitions of treatment and control groups**

| | Broad- | DHS | | | | | Afrobarometer | | | | |
|---|---|---|---|---|---|---|---|---|---|---|---|
| | band | Treatment | | Control | | | Treatment | | Control | | |
| Specification | proxy | Rate | $N$ | Rate | $N$ | $t$ | Rate | $N$ | Rate | $N$ | $t$ |
| Original, 0.005°/ 0.1° from lines | GSMA | 0.67 | 3,414 | 0.57 | 17,162 | 1.83 | 0.52 | 1,257 | 0.69 | 3,007 | –4.00 |
| | Ookla | 0.43 | 5,653 | 0.22 | 30,869 | 4.90 | 0.32 | 1,622 | 0.39 | 4,108 | –2.11 |
| + Re-geocode, recalculate distance | GSMA | 0.46 | 7,740 | 0.30 | 40,593 | 4.26 | 0.77 | 1,119 | 0.75 | 3,088 | 0.45 |
| | Ookla | 0.32 | 12,837 | 0.15 | 68,082 | 6.98 | 0.50 | 1,503 | 0.38 | 4,240 | 3.01 |
| + Change to 0.05°/ 0.1° from lines | GSMA | 0.39 | 36,725 | 0.12 | 11,608 | 11.38 | 0.77 | 3,718 | 0.67 | 489 | 1.90 |
| | Ookla | 0.22 | 62,693 | 0.04 | 18,226 | 14.00 | 0.45 | 5,075 | 0.15 | 668 | 8.74 |
| + Double to 0.1°/0.2° from lines | GSMA | 0.33 | 48,333 | 0.10 | 17,255 | 10.82 | 0.76 | 4,207 | 0.28 | 635 | 10.38 |
| | Ookla | 0.18 | 80,919 | 0.01 | 25,295 | 16.99 | 0.41 | 5,743 | 0.08 | 994 | 12.86 |
| + Change to 0.1°/0.2° from node | GSMA | 0.50 | 27,123 | 0.20 | 20,870 | 10.80 | 0.86 | 2,900 | 0.62 | 1,047 | 5.63 |
| | Ookla | 0.24 | 52,406 | 0.05 | 30,777 | 12.55 | 0.46 | 4,105 | 0.28 | 1,603 | 5.15 |
| | | South Africa QLFS | | | | | Ethiopia LMMIS | | | | |
| Original, 0.005°/ 0.1° from lines | GSMA | 0.97 | 24,168 | 0.91 | 155477 | 5.89 | | | | | |
| | Ookla | 0.31 | 24,168 | 0.21 | 155477 | 3.42 | 0.06 | 2,496 | 0.11 | 453 | –0.44 |
| + Re-geocode, recalculate distance | GSMA | 0.97 | 25,470 | 0.90 | 157222 | 6.53 | | | | | |
| | Ookla | 0.30 | 25,470 | 0.21 | 157222 | 3.38 | 0.08 | 2,758 | 0.03 | 412 | 0.74 |
| + Change to 0.05°/ 0.1° from lines | GSMA | 0.94 | 147,938 | 0.81 | 34,754 | 6.08 | | | | | |
| | Ookla | 0.25 | 147,938 | 0.11 | 34,754 | 8.15 | 0.08 | 3,139 | 0.06 | 31 | 0.12 |
| + Double to 0.1°/0.2° from lines | GSMA | 0.91 | 182,692 | 0.67 | 22,693 | 8.06 | | | | | |
| | Ookla | 0.22 | 182,692 | 0.04 | 22,693 | 12.61 | 0.08 | 3,170 | 0.00 | 4 | 1.35 |
| + Change to 0.1°/0.2° from node | GSMA | 0.93 | 122,228 | 0.82 | 63,847 | 7.18 | | | | | |
| | Ookla | 0.25 | 122,228 | 0.15 | 63,847 | 5.73 | 0.03 | 2,738 | 0.44 | 178 | –1.69 |

Notes: The table shows the fraction of various data sets (not) covered by broadband, restricting to observations that contribute identification in later regressions. Two proxies for broadband are used: "strong" 3G coverage as of 2014, according to data from the GSMA, which covers Ghana, Kenya, Madagascar, Nigeria, South Africa, and Tanzania; and Ookla-measured download speeds in the first quarter of 2019. $t$ statistics are for the hypothesis of no difference in coverage between treatment and control groups, and indicate the fidelity of a treatment/control delineation to actual broadband access. In the first pair of rows in each subtable, variables and samples are defined as in HP19; in particular, outer treatment and control radii are 0.005° and 0.1°. Subsequent pairs accumulate changes. In the second pair, variables are revised as described in the notes to Table 3 and Table 4. In the third, the treatment radius is increased to 0.05°. In the fourth, radii are doubled to 0.1° and 0.2°. In the last, network nodes active in 2010 are the distance referent. Standard errors clustered by locality are in parenthesis.

# 3 Results from non-satellite data

## 3.1 Geocoding

Recall that the Akamai, Ethiopia LMMIS, and Afrobarometer data sets are here

recoded from place names in the primary sources—the last through recourse to the AidData effort (BenYishay et al. 2017). For Akamai and Afrobarometer, I also have the HP19 longitude and latitude estimates. Figure A–2 and Figure A–3 in the appendix therefore compare the old and new coordinate assignments by drawing lines between corresponding locations and marking the new ones with dots. Seven of the nine Afrobarometer countries have at least one locale that HP19 places outside the nation's borders.

To better understand the disagreements, I closely examine the old and new geocodings within manageable subsamples. For the Akamai data set, I narrow to the 142 locations with at least 10 download speed readings (an HP19 inclusion criterion) that are within 0.1° of fiber under at least one of the geocodings. Among the 20 largest disagreements, the main issue is that some Akamai city names are ambiguous. For example, at least four villages in Nigeria are named "Asa."[15] It appears that HP19 erroneously *averages* the coordinates of all places with the same name, and even places whose names begin the same way. The major South African city of Durban, which is on the country's east coast, is geolocated by HP19 to the middle of the country, halfway between its true location and the western town of Durbanville. A few other "cities" are states or provinces; here they are left uncoded because of their geographic coarseness.[16]

Similarly, for Afrobarometer I carry out a detailed comparison of the HP19 and AidData coordinates in Benin, which is the alphabetically first country in the sample.[17] This reveals kilometer-scale errors in geocodings. However, the HP19 geocoding also contains larger errors. One cause is a bug that gives the same coordinates to different locations of the same name within the same country, such as “1er ARROND” (first arrondissement) in the cities of Cotonou, Parakou, and Porto-Novo. When restricting, as here, to the locations AidData marks as highest-precision, I estimate the average AidData error in Benin at 0.026° (2.84km). The average error in HP19 is 0.22° (24.6km).

[15] See nominatim.openstreetmap.org/search?q=asa&countrycodes=NG.

[16] I also revisit the list of top-four largest cities in each country, which HP19 exclude from some Akamai regressions because of ambiguity in their treatment classification. I add Porto-Novo, Benin; Kisangani, DRC; Tema, Ghana; Makuru, Kenya; Nampula, Mozambique; and Dodoma and Zanzibar, Tanzania. I drop Arusha, Tanzania; and Bloemfontein and East London, South Africa. The source for all changes is, as in HP19, UNdata (data.un.org/Data.aspx?d=POP&f=tableCode%3a240, using pre-2015 observations) except for the DRC, which is absent from that source, and for which GeoNames is used (geonames.org/CD/largest-cities-in-.html).

To assess the impact of the geocoding revisions on treatment classification of the Afrobarometer observations, Figure 3 plots each survey locale at its AidData location, using symbols to indicate agreement or disagreement between old and new codings under HP19’s preferred 0.005°/0.1°-from-lines classification. Figure A–4 through Figure A–7 in the appendix do the same for the Akamai, DHS, South Africa QLFS, and Ethiopia LMMIS data sets.[18] Discrepancies are rare for the DHS and QLFS, presumably because these geocodings allow the least researcher discretion: the DHS locations come directly from the data provider while the QLFS locations are centroids of polygons provided by Statistics South Africa. The only source of disagreement, is that the new distances are computed in ellipsoidal rather than planar geometry. Overall, the old and new treatment-control classifications agree on 68.6% of locales for Akamai, 99.7% for DHS,

[17] I start by reverse geocoding from the GADM database for Benin (gadm.org/download_country.html), which provides boundaries of third-level administrative units, i.e., arrondissements. If HP19 and AidData “GADM-agree”—if their coordinates land in the same arrondissement according to GADM—I average the two sources’ coordinates. If only one source is “GADM-correct,” I copy its coordinates. Otherwise I use the geographic centroid of the named arrondissement according to GADM. But, overriding the foregoing, if the Afrobarometer place label identifies a village or neighborhood *within* an arrondissement, as is common, I attempt to locate the place using online sources, and usually succeed. Sources are documented through hyperlinks in the Excel table.

[18] The HP19 versions of these data sets that are available to me do not include geographic coordinates, but do include a distance variable sufficient for assigning treatment status.

70.7% for Afrobarometer, 98.1% for the South Africa QLFS, and 83.5% for the Ethiopia LMMIS.[19]

**Figure 3. Treatment classification comparison, Afrobarometer household survey locations**

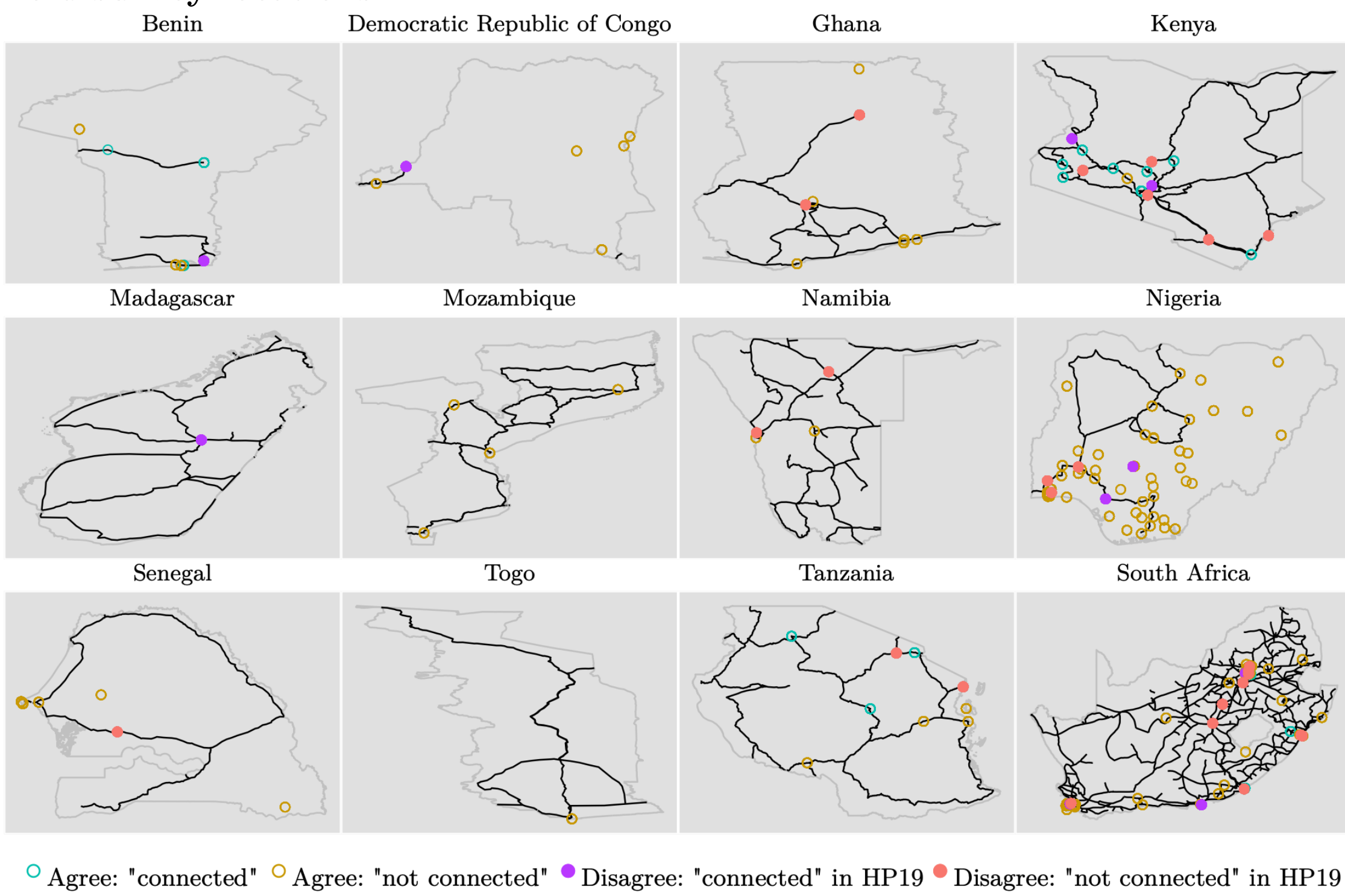

Notes: The HP19 network maps are drawn in black, and national borders in grey. Hollow circles show Afrobarometer enumeration areas where the HP19 and new treatment classifications agree on treatment assignment under the HP19 classification (within 0.005° of a fiber line for treatment). Solid circles show disagreements. Circles are plotted at their AidData coordinates (BenYishay et al. 2017). Scales are chosen separately for each axis of each plot.

## 3.2 Replication and revision: Internet speed and usage

The first impact estimates in HP19 are for Internet speed and the frequency of Internet use. The speed data come from Akamai and the usage data from Afrobarometer. In the language of instrumental variables, these regressions can be viewed as the first stage in HP19's assessment of the impact of submarine cables. They assess the association between the variable that is the primary source of identifying variation and more proximate indicators of treatment.

[19] "Classification" here is trichotomous: treatment, control, or excluded.

Unfortunately, the data analyzed in this first stage is among the most coarsely geocoded in the study. As a result, it appears, the estimates are fragile. The Akamai Internet speed data are coded only by city (or province or state in some cases). And for this first-stage analysis, HP19 replaces Afrobarometer survey locations with the nearest Akamai cities. Notably, precisely because in the WBES firm survey locations are also only coded by city, HP19 deemphasizes results from that data set, reporting them in an appendix.[20]

Here, because the details of the mapping between Afrobarometer and Akamai locations are absent from the HP19 code archive, I reimplement the Internet usage regressions by copying HP19's Afrobarometer employment specifications. This change has the additional benefits of constituting a minimally arbitrary robustness test and more directly examining whether, within the analytical frame for estimating reduced-form employment impacts, first-stage effects can be detected.

The first three columns of the first row of Table 2 replicate HP19's estimates of the impact of submarine cable connection on Akamai-measured bandwidth. The result in the upper left corner repeats the original finding that undersea connection increased bandwidth by 0.35 arc-hyperbolic-sine (asinh) points. That specification controls for country-year and location fixed effects and excludes location-quarters with 10 or fewer speed measurements. Following HP19, column 2 introduces an exclusion for large cities, since they ambiguously encompass neighborhoods within and beyond the reach of broadband. Also following HP19,

[20] A more precise reading of HP19 may be that it deemphasizes the WBES results for two reasons: "In light of the WBES firms being clustered in space *and* the approximate connectivity classification required due to the relatively aggregate geographical level at which their location is reported… [emphasis added]." However, all the survey data in HP19 comes from complex surveys that cluster the sampling, so the first reason does not seem like an operative distinction.

column 3 moves to the triple-difference specification, adding treatment group–year dummies.

The final columns of the first row present the original impact estimates for Afrobarometer-reported Internet usage by individuals—in particular, for whether people log on at least daily or at least weekly. HP19's double-differenced estimates are included while the triple-differenced ones are dropped since they are not actually identified (see section 1).

Much as in the previous table, the remaining rows of Table 2 accumulate specification changes. In the second row, HP19's definitions of the treatment and control zones are retained. But when working with the Akamai data whose cadence is quarterly, quarterly dummies replace yearly ones, in conformity with the TWFE design and the precedent set by HP19's regressions on South Africa QLFS data. The list of large cities is revised, as described in footnote 16. For both Akamai and Afrobarometer, the revised geocodings are introduced. And the switch is made to the Afrobarameter specification that HP19 uses to estimate impacts on employment. The point estimates all fall, becoming indistinguishable from 0, or even negative. The third and fourth rows switch to the preferred treatment definitions, using radii of 0.1° and 0.2° from fiber lines or nodes. The estimates are usually negative, albeit without much statistical significance.

These results stand in tension with the clear treatment-control contrasts drawn in Table 1 when using 0.1°/0.2° bounds. How are they squared? The earlier results are (post-treatment) differences, whereas those in Table 2 are differences in differences. The Table 1 results are for broadband proxies with a resolution of about 600 meters (for Ookla speeds) or 260 meters (for 3G coverage) whereas the Table 2 results reference Akamai download data points geolocated only to the city. As a result, while there is little doubt that proximity to fiber is associated with broadband access, and little doubt that undersea cable connection circa 2010 increased broadband availability in Africa, it is less clear that it did so *differentially* with respect to distance from fiber, at least enough so to generate a detectable signal in the specifications and data brought to bear here.

**Table 2. Impact of submarine cable arrival on Internet speed and usage**

| | Akamai | | | Afrobarometer | |
|---|---|---|---|---|---|
| Specification | Log bandwidth | | | Use Internet daily | Use Internet weekly |
| Original, 0.005°/ 0.1° from lines | 0.354 | 0.362 | 0.380 | 0.082 | 0.123 |
| | (0.134) | (0.171) | (0.185) | (0.028) | (0.032) |
| Observations | 2,533 | 1,670 | 1,670 | 4,160 | 4,160 |
| + Re-geocode, recalculate distance | –0.043 | 0.017 | 0.154 | –0.035 | –0.007 |
| | (0.142) | (0.195) | (0.231) | (0.021) | (0.030) |
| Observations | 2,661 | 1,660 | 1,660 | 10,353 | 10,363 |
| + Change to 0.1°/0.2° from lines | –0.401 | –0.249 | –0.535 | 0.011 | –0.038 |
| | (0.299) | (0.441) | (0.203) | (0.025) | (0.027) |
| Observations | 2,734 | 1,706 | 1,700 | 12,190 | 12,200 |
| + Change to 0.1°/0.2° from nodes | –0.156 | –0.292 | –0.053 | –0.023 | –0.036 |
| | (0.286) | (0.336) | (0.357) | (0.023) | (0.030) |
| Observations | 2,845 | 1,827 | 1,827 | 10,663 | 10,671 |
| Treatment zone–time FE | No | No | Yes | | |
| Include biggest cities | Yes | No | No | | |

Note: Each cell holds an estimate of a coefficient on the treatment term, a dummy for being close to national fiber network after it has been connected to an undersea cable circa 2010. The first row is copied from HP19 Table 2. The treatment group is within 0.005° of a fiber line and the control zone within 0.005–0.1°. The Afrobarometer estimates in columns 5 and 7 of HP19 Table 2 are not included because those estimates are not identified (see section 1 above). Subsequent rows accumulate changes toward the preferred specifications at bottom. In the second row, observations are re-geocoded, the list of four biggest cities in each country is revised, quarterly dummies replace yearly ones in the Akamai regressions, fixed-effect singletons are dropped, and the Afrobarometer specifications are switched to match the HP19 employment regressions on Afrobarometer data. For the Internet usage results, a seemingly erroneous HP19 sample restriction is also dropped; see notes to Table 4. In the third row, bounds for the treatment and non-treatment groups are extended to 0.1° and 0.2°. In the last, fiber optic nodes become the referent for the geographic dimension of the treatment definition. All regressions include locality dummies. Standard errors clustered by locality are in parenthesis.

## 3.3 Replication and revision: Employment

HP19's primary outcome is employment, at the level of the individual or the firm. Impacts are estimated for overall employment and, as data allow, by the skill level of the work and the education level of the worker. Table 3, Table 4, and Table 5 revisit the individual-level employment results from the DHS, Afrobarometer, and South Africa QLFS data sets. Table 6 does the same for firm-level employment in the Ethiopia LMMIS. All follow Table 2 in format.

**DHS**

The first row of Table 3 recapitulates the HP19 employment results from DHS data: submarine connection is followed by a 4.6 percentage point increase in employment, with the effect concentrated in skilled occupations but spread rather evenly across all but the least-educated group. All the underlying regressions include controls for country-time effects and grid cell–treatment status effects, where grid cells are 0.1°×0.1°. Standard errors are clustered by grid cell. The samples cover eight countries, with one pre- and one post-treatment survey from each.

The second major row sticks to the HP19 0.005°/0.1°-from-lines classification but introduces corrections. Distances from fiber are recomputed in ellipsoidal geometry. An exclusion in HP19 of observations whose interview year differs from the nominal survey year is reversed; this matters for DHS surveys that interviewed people across two calendar years. And some respondents who are marked in HP19 as having skilled or unskilled employment, yet also as not being employed, are now excluded from the regressions by skill level. The changes lower the overall employment impact by a third and remove the impression of skill bias. The pattern of results across education levels does not change much.

Drawing treatment and control boundaries at 0.1° and 0.2°, whether from lines or nodes, further reduces the apparent impact on overall employment and on employment of the most educated (last two major rows of Table 3). To complement the geographic coarsening of the treatment and comparator definitions, grid cells are now enlarged to 0.2°×0.2°, as in Hjort and Poulsen (2016). The overall employment estimate (first column) is 1.8 points when taking distance from lines (standard error 1.4) and 1.7 points (1.3), when taking distance from nodes. To this extent, the HP19 finding of an employment impact in the DHS data is preserved. The finding of skill bias looks less robust.

**Table 3. Impact of submarine cable arrival on individual employment (DHS)**

| Specification | Any employment | Skill level of job: Skilled | Skill level of job: Unskilled | Education level: Primary not completed | Education level: Primary | Education level: Secondary | Education level: Higher |
|---|---|---|---|---|---|---|---|
| Original, 0.005°/ 0.1° from lines | 0.046 | 0.044 | 0.003 | –0.013 | 0.061 | 0.047 | 0.067 |
| | (0.014) | (0.018) | (0.015) | (0.032) | (0.022) | (0.015) | (0.027) |
| Observations | 59,914 | 59,966 | 59,923 | 59,912 | 59,912 | 59,912 | 59,912 |
| + Revise sample, recalculate distance | 0.031 | 0.013 | 0.017 | –0.019 | 0.038 | 0.032 | 0.047 |
| | (0.012) | (0.014) | (0.011) | (0.021) | (0.017) | (0.013) | (0.022) |
| Observations | 121,969 | 121,969 | 121,969 | 121,965 | 121,965 | 121,965 | 121,965 |
| + Change to 0.1°/0.2° from lines | 0.018 | 0.006 | 0.012 | –0.002 | 0.033 | 0.017 | –0.011 |
| | (0.014) | (0.015) | (0.016) | (0.018) | (0.015) | (0.014) | (0.019) |
| Observations | 157,358 | 157,358 | 157,358 | 157,352 | 157,352 | 157,352 | 157,352 |
| + Change to 0.1°/0.2° from nodes | 0.017 | –0.018 | 0.035 | 0.000 | 0.033 | 0.020 | –0.018 |
| | (0.013) | (0.013) | (0.014) | (0.020) | (0.015) | (0.014) | (0.018) |
| Observations | 131,022 | 131,022 | 131,022 | 131,018 | 131,018 | 131,018 | 131,018 |

Note: Results in first row exactly or closely match those in HP19 Table 3, panel A, column 1; Table 5, panel A, columns 1 and 2; and Table 6, first panel, column 1. For them, treated units lie within 0.005° of a fiber line and comparators within 0.005–0.1°. Subsequent rows accumulate changes. In the second, an erroneous exclusion of observations with interview year different from nominal survey year is reversed; distances are computed in ellipsoidal rather than planar geometry; and observations marked as having no employment but having skilled employment are treated as unemployed throughout. In the third, treatment and comparator radii are extended to 0.1° and 0.2°. In the last, distance is taken from the nearest network node rather than network line, using 2010 node locations from Africa Bandwidth Maps. In each row, results in the last four columns are coefficients on interaction terms within the same regression. All regressions include dummies for country-year and grid cell–treatment group combinations. Grid cells are 0.1°×0.1° in the first two rows, as in HP19, and 0.2°×0.2° in the last two, as in Hjort and Poulsen (2016). Standard errors clustered by grid cell are in parenthesis.

### Afrobarometer

The first row of the Afrobarometer impacts table (Table 4) replicates the HP19 estimates for total employment and employment by education level. In the second row, a handful of observations with interview year differing from nominal survey year are restored.[21] The new geocoding is introduced, and distances are recomputed as before. These revisions erase the apparent impact on employment in this data set. Widening the treatment and control zones, as reported in the third and fourth major rows, also leads to point estimates hard to distinguish from zero.

[21] Previous versions of this comment also departed from HP19 in including in the Afrobarometer regressions those who are neither working nor looking for work—on the misunderstanding that the HP19 DHS and South Africa QLFS regressions include them too. In fact, only the DHS regressions do, because of data limitations.

**Table 4. Impact of submarine cable arrival on individual employment (Afrobarometer)**

| | | Education level | | | |
|---|---|---|---|---|---|
| Specification | Any employment | Primary not completed | Primary | Secondary | Higher |
| Original, 0.005°/ 0.1° from lines | 0.077 | 0.110 | 0.078 | 0.061 | 0.098 |
| | (0.036) | (0.054) | (0.040) | (0.048) | (0.050) |
| Observations | 7,918 | 7,902 | 7,902 | 7,902 | 7,902 |
| + Revise sample, re-geocode, recalculate distance | –0.001 | –0.030 | –0.029 | –0.003 | –0.007 |
| | (0.036) | (0.061) | (0.042) | (0.041) | (0.050) |
| Observations | 7,725 | 7,711 | 7,711 | 7,711 | 7,711 |
| + Change to 0.1°/0.2° from lines | 0.049 | 0.061 | 0.031 | 0.064 | 0.036 |
| | (0.053) | (0.054) | (0.050) | (0.052) | (0.054) |
| Observations | 9,021 | 9,007 | 9,007 | 9,007 | 9,007 |
| + Change to 0.1°/0.2° from nodes | –0.045 | 0.015 | –0.064 | –0.041 | –0.073 |
| | (0.036) | (0.044) | (0.039) | (0.041) | (0.045) |
| Observations | 7,861 | 7,848 | 7,848 | 7,848 | 7,848 |

Note: This table has the same format as Table 3 except for lacking columns for skilled and unskilled employment. See notes to that table. Results in first row exactly or closely match those in HP19 Table 3, panel A, column 2, and Table 6, second panel, column 1. Starting in the second row, the AidData geocoding of all locations is used and distances are computed in ellipsoidal rather than planar geometry. Standard errors, clustered by 0.1°×0.1° grid cell in the first two rows and 0.2°×0.2° in the last two, are in parenthesis.

### South Africa QLFS

Results for the South Africa labor force survey appear in Table 5. Following HP19, enumeration areas replace grid cells as the basis for spatial fixed effects and error clustering. Recomputing distances in ellipsoidal geometry (second row) modestly reduces the point estimate of total employment impact, from 2.2 to 1.7 points. The estimate falls to 0.006 or 0.000 in the wider-zone specifications. And, as in the DHS results, there is now no clear bias for or against skilled employment.

**Table 5. Impact of submarine cable arrival on individual employment (South Africa QLFS)**

| Specification | Any employment | Skilled | Unskilled | Primary not completed | Primary | Secondary | Higher |
|---|---|---|---|---|---|---|---|
| Original, 0.005°/ | 0.022 | 0.014 | –0.001 | 0.012 | 0.028 | 0.022 | 0.019 |
| 0.1° from lines | (0.008) | (0.006) | (0.005) | (0.017) | (0.011) | (0.012) | (0.011) |
| Observations | 280,641 | 280,641 | 280,641 | 277,737 | 277,737 | 277,737 | 277,737 |
| + Recalculate | 0.017 | 0.009 | 0.009 | 0.010 | 0.023 | 0.016 | 0.015 |
| distance | (0.008) | (0.009) | (0.007) | (0.016) | (0.011) | (0.011) | (0.011) |
| Observations | 284,063 | 284,063 | 284,063 | 281,132 | 281,132 | 281,132 | 281,132 |
| + Change to 0.1°/0.2° | 0.006 | 0.008 | –0.002 | –0.001 | –0.003 | 0.007 | 0.015 |
| from lines | (0.011) | (0.011) | (0.008) | (0.013) | (0.012) | (0.012) | (0.012) |
| Observations | 312,394 | 312,394 | 312,394 | 309,167 | 309,167 | 309,167 | 309,167 |
| + Change to 0.1°/0.2° | 0.000 | 0.008 | –0.008 | –0.005 | –0.006 | 0.004 | 0.004 |
| from nodes | (0.007) | (0.006) | (0.006) | (0.010) | (0.007) | (0.008) | (0.008) |
| Observations | 285,735 | 285,735 | 285,735 | 282,763 | 282,763 | 282,763 | 282,763 |

Note: This table has the same format as Table 3, which see. Results in first row are from HP19 Table 3, panel A, column 3; Table 5, panel B, columns 1 and 2; and Table 6, third panel, column 1. Unlike in the previous tables, grid cell fixed effects are not controlled for, and all standard errors are clustered by enumeration area. The skilled and unskilled results in the first row are copied directly from HP19 because they could not be replicated with the public data and code.

### Ethiopia LMMIS

HP19 finds a large effect on firm-level employment in Ethiopia, at 0.156 or 0.224 log points depending on the choice of fixed effects. Here too, skilled employment appears to increase more. Those results appear in the top row of Table 6. The revision to the geocoding of Ethiopia LMMIS observations documented in Figure A–7, and the recomputation of the distances from the fiber network, do not appreciably change these estimates (second row). However, as in South Africa, revising the treatment and control zones greatly weakens the results. Then, almost none are easily distinguishable from zero.

**Table 6. Impact of submarine cable arrival on firm employment (Ethiopia LMMIS)**

| Specification | Employees | | Skilled employees | | Unskilled employees | |
|---|---|---|---|---|---|---|
| Original, 0.005°/ | 0.156 | 0.224 | 0.034 | 0.201 | 0.115 | 0.124 |
| 0.1° from lines | (0.091) | (0.081) | (0.080) | (0.106) | (0.078) | (0.095) |
| Observations | 5,360 | 5,360 | 5,360 | 5,360 | 5,360 | 5,360 |
| + Re-geocode, recalculate | 0.151 | 0.197 | 0.044 | 0.222 | 0.259 | 0.172 |
| distance | (0.083) | (0.067) | (0.105) | (0.095) | (0.063) | (0.087) |
| Observations | 5,737 | 5,696 | 5,737 | 5,696 | 5,737 | 5,696 |
| + Change to 0.1°/0.2° from | –0.092 | 0.018 | –0.055 | –0.334 | –0.309 | –0.065 |
| lines | (0.123) | (0.134) | (0.267) | (0.378) | (0.106) | (0.099) |
| Observations | 5,748 | 5,707 | 5,748 | 5,707 | 5,748 | 5,707 |
| + Change to 0.1°/0.2° from | –0.042 | 0.040 | –0.082 | 0.017 | –0.191 | –0.058 |
| nodes | (0.105) | (0.148) | (0.144) | (0.179) | (0.108) | (0.166) |
| Observations | 5,269 | 5,228 | 5,269 | 5,228 | 5,269 | 5,228 |
| Year FE | Yes | No | Yes | No | Yes | No |
| Treatment zone–grid cell FE | No | Yes | No | Yes | No | Yes |
| Industry-year FE | No | Yes | No | Yes | No | Yes |
| Firm FE | Yes | No | Yes | No | Yes | No |

Note: This table has the same format as Table 3, which see. Results in first row exactly or closely match those in HP19 Table 8, first row. Dependent variables are subject to the arc-hyperbolic sine transform. Standard errors, clustered by 0.1°×0.1° grid cell in the first two rows and 0.2°×0.2° in the last two, are in parenthesis.

**World Bank Enterprise Surveys**

HP19 runs similar specifications on the firm-level data from the multi-country WBES in order to study impacts on firm-level exports, on-the-job training, and corporate use of the Internet. The regressions indicate a 14–17 log point increase in firm employment after undersea cable connection. But, as noted above, HP19 places less confidence in the results because survey locations are only coded by city.

HP19's deemphasis of the WBES results is even more warranted than it appears at first. All of the identification flows from Nigeria data, where the geographic coding is especially coarse, and thus the treatment assignment is especially uncertain. Concretely, if Nigeria is dropped from the WBES regressions, the treatment indicator becomes collinear with the fixed-effect dummies. Why? In the HP19 subset of the WBES data, Nigeria is the only country to have more than four locales visited in any one round. In each of the other countries, those cities that received surveyors both before and after submarine cable connection were either all deemed close to fiber lines in the HP19 coding, or all far. Only the Nigeria data can fill all the boxes in the classical 2×2 difference-in-differences grid, and therefore contribute identifying information under HP19's pooled, multi-country TWFE design. And for Nigeria, the "firm establishment location" field turns out to contain names of states, not cities. Since firms cannot be geolocated with much precision when only their home states are known, I fully set aside the WBES analysis.

## 3.4 Varying the treatment radius more systematically

While HP19 prefers a treatment radius of 0.005° (~560 meters), the study performs a robustness check on that choice. For the individual-level employment impact estimates, Figure 4 of HP19 varies the radius between 400 and 2,000 meters for DHS and South Africa QLFS and between 400 and 5,000 meters for Afrobarometer. To address the tension between the arguable robustness in that figure and the greater fragility reported here, I replicate and extend the HP19 figure. I add the Akamai and Ethiopia regressions to the testing, and bring Afrobarometer's radius range, 400–5,000 meters, to all datasets. I also incorporate the revisions embodied in the second row of the above tables, for example, including observations whose interview year differs from the nominal survey year. And for the multi-country specifications, I correct one error in the HP19 code: while the tabulated HP19 regression results incorporate country–time fixed effects, the Figure 4 regressions incorporate only time effects. The new results are plotted here in Figure 4.

In the data sets where geocoding is revisited because the primary data contains only place names—Akamai, Afrobarometer, and Ethiopia LMMIS, the results change substantially. The Akamai coefficient is hard to distinguish from zero in both the original and revised specifications. The Afrobarometer coefficient is much lower in the revised specification, while the Ethiopia coefficient is higher. Results are more stable in the DHS and South Africa QLFS data sets. In the DHS, the employment impact hovers around 2 percentage points, which coheres with the preferred DHS estimates in the lower half of Table 3 above. In the South Africa data, the estimate moves to zero as the treatment radius extends beyond 2,000 meters. Since HP19 does not provide a specific technological basis for the 0.005° radius, and since the evidence in Table 1 favors longer radii, it is reasonable to put less interpretive weight on the results at the left edges of these plots, though they correspond to HP19's preferred estimates.

**Figure 4. Estimated effect of submarine cable connection on internet speed and employment, varying the treatment radius**

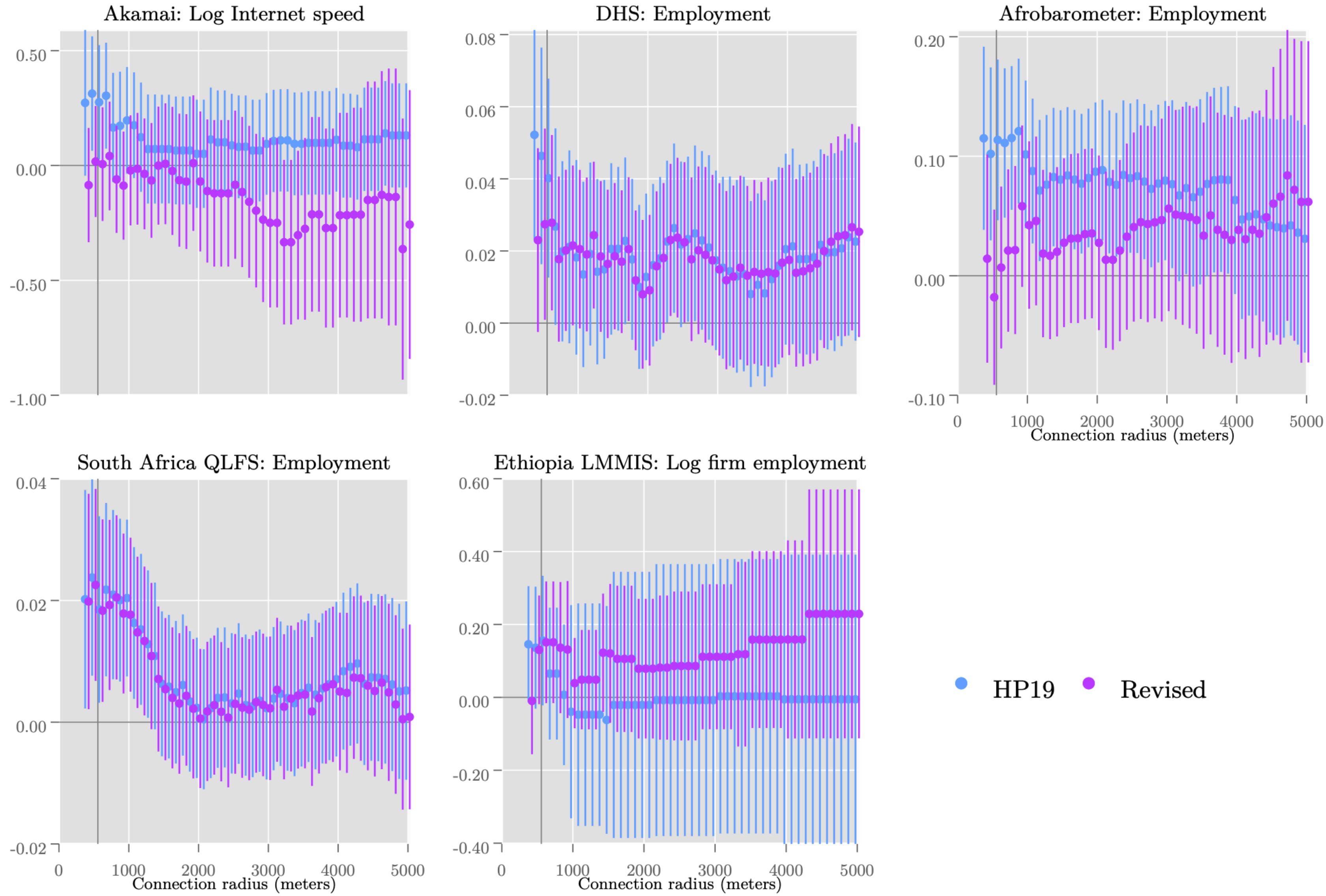

Notes: Following HP19, Figure 4, each plot shows how the results change as the outer bound for the treatment zone is varied between 400 and 5,000 meters. The base specifications are in the first columns, first and second rows, of Table 2–Table 6.

### 3.5 Meta-analysis

To distill the results on individual-level employment from the DHS, Afrobarometer, and South Africa QLFS data sets, I perform meta-analysis. The spirit of the exercise is to average the estimates in a way that factors in their relative precision and minimizes researcher discretion.[22] The default method in Stata's "meta forestplot" command, DerSimonian-Laird random-effects meta-analysis, is applied. This method does not assume that the average treatment effect is homogeneous across settings, and here must estimate the heterogeneity with a sample of only three "studies." See Figure 5. The meta-analytically averaged estimate of the impact of broadband on employment is 3.6 percentage points for the HP19 data and specifications, with a standard error of 1.2 points (first pane). The estimate drops to 2.0 points (standard error 0.7) with data, specification, and geocoding revisions (second pane), and to 1.2 (standard error 0.9; third pane) with the wider treatment and control zones. Although the forest plot emphasizes that this estimate is not significant at $p = 0.05$, it nearly is; the confidence interval lies almost entirely in positive territory. Last, classifying with 0.1°/0.2°-from-nodes yields a meta-analytical point estimate of 0.3 points (standard error 0.6).

The main source of positivity in the revised results is the DHS data, where the lag between pre- and post-treatment measurement is longest (9 years), and where, as a result, the exogenous shocks of submarine cable connection account for the smallest share of the temporal component of identifying variation.

[22] Otis Reid suggested this step.

**Figure 5. Meta-analytical summary: fast Internet and employment**

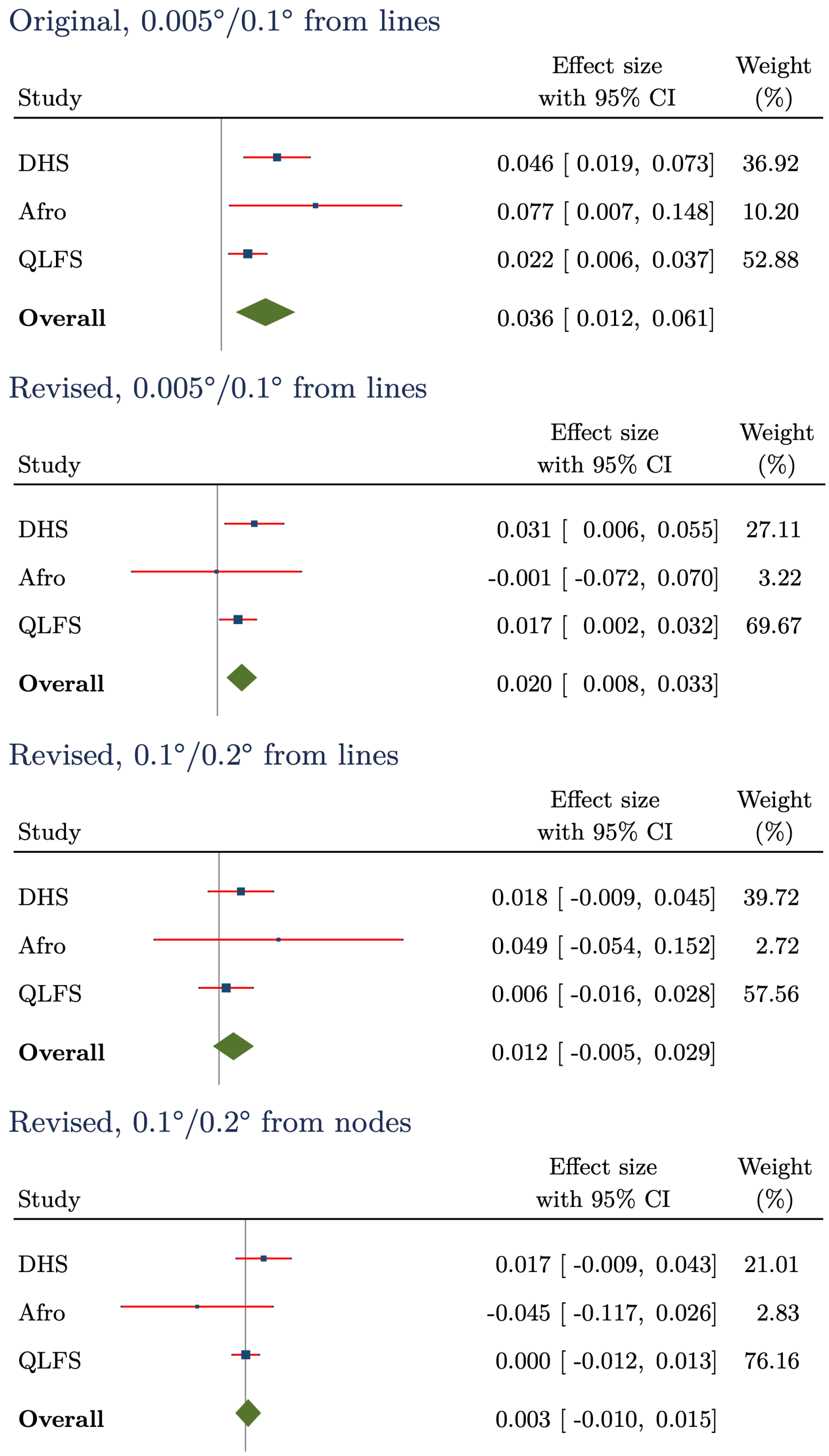

Notes: Using DerSimonian and Laird random-effects meta-analysis, these forest plots combine employment impact estimates from the first columns of Table 3, Table 4, and Table 5.

## 4 Results for nighttime light

Analysis of satellite data is a promising way to quantify economic development (Elvidge et al. 1997; Chen and Nordhaus 2011; Henderson, Storeygard, and Weil 2012). But there are pitfalls, especially when using data from the early-generation DMSP-OLS satellite system. That system was designed not for social science but for measuring cloud cover to help the U.S. Air Force forecast the weather (Gibson, Olivia, and Boe-Gibson 2020).

One major complication in the "stable lights" DMSP data sets used in economics is that the values are not strictly comparable over time. The Air Force regularly adjusted the satellite sensors, notably to compensate for the waxing and waning of the moon. The specifics of this variation are not documented. Moreover, the stable light values are annual averages at each location over cloud-free days. Since cloud cover variability could intersect with the lunar cycle in complex ways, the amplification baked into each pixel varies over space and time. A final complication is that a major disruption occurred within the HP19 timeframe: the DMSP switched satellites between 2009 and 2010, from satellite F16 to F18.[23] At that juncture, the global sum of the light data jumped by more than half—by far the largest year-to-year change (Li and Zhou 2017, Figure 1).

These problems are well recognized, at least outside of economics (Gibson, Olivia, and Boe-Gibson 2020). And perhaps they are why, within economics, researchers tend to use the DMSP data in ways that minimize the influence of short-term changes, working with pure cross-sections (Bleakley and Lin 2012, Michalopoulos and Papaioannou 2013) or long differences.[24]

A common response to the incomparability over time is to "intercalibrate" different years' data using the observations of *Sicily* (Elvidge et al. 2009). Since

[23] Earth Observation Group, eogdata.mines.edu/products/dmsp/#download, accessed April 8, 2023.

[24] "Most empirical studies using light data have focused on cross-section variation." Lowe (2014, p.10).

Sicily is part of a mature economy, its light emissions might not change much year-to-year. And its pixel-level readings are relatively evenly distributed across the measured power range. Typically, to calibrate against Sicily, the coefficients in a quadratic transformation are tuned to minimize the least-squares discrepancy between the benchmark year's readings for Sicily and a given year's intercalibrated values. The quadratic transformation is then applied globally.

HP19 takes an alternative approach: forgoing calibration and relying on year dummies to expunge systematic shifts over time, as in Chen and Nordhaus (2011) and Henderson, Storeygard, and Weil (2012).

Neither approach works perfectly. As just explained, whatever causes the gain shifts does not affect all pixels equally. What's best for Sicily is not best for Africa. And unremoved artifacts that are second-order when taking first differences, such as when comparing regions, can become first-order in second differences. For its part, the fixed-effect approach only works to the extent that the gain shifts respect the transformations that the researcher applies to the light data. In the best case, which is unrealistic, researchers take the light values in logarithms, and calibrating each year's data only requires multiplication by some global constant. Then, the logarithmic gain will be perfectly absorbed by year fixed effects. But if that were sufficient, intercalibration needn't involve a nonlinear transform (e.g., quadratic). And in HP19, the light values enter regressions after passing through the arc-hyperbolic sine (asinh) function. It would be surprising if, after applying this transform, the time effects are constant across the power spectrum, thus fully expungable by fixed-effect controls.

HP19 runs two nighttime light regressions, one double-differenced and one triple-differenced. To demonstrate the potential for confounding from the satellite switch, I adapt the first one into an event study. The single treatment-dummy—for being in the treated zone in the post-treatment period—is replaced by its interactions with year dummies, taking 2013 as the reference year. HP19's 0.005°/0.1°-from-lines treatment definition is copied. But the original, high-resolution satellite data are used, and an HP19 bug causing the post-treatment dummy to sometimes exceed 1 is fixed. The event study is run on the raw light data as in HP19, and on intercalibrated data. See Figure 6. In the raw data, the effect of being in the treatment zone, relative to the control zone, jumps dramatically just when the DMSP data switches from satellite F16 to F18. In the intercalibrated data, the jump may still be present but is not as clean: it is plausible that intercalibration removes most of the effect of the satellite change, and plausible that it does not.

HP19 also reports an event study (Figure A3). There, the event is undersea cable landing rather than the satellite change, so that relative to the event study shown here, some countries' information is shifted forward or back in time. Under the null that confounding from the satellite switch fully explains the HP19 results, the HP19 event study would look very much as it as actually does. The universal, sharp rise between 2009 and 2010 would be smoothed by the partial shifting into a gradual climb.

To remove the confounding, I will introduce a control into the HP19 double-difference specification: the product of the dummies for being in the treated zone and being observed post-2009. It will absorb any systematic tendency for the treatment-control difference in luminosity to shift as a result of the changeover from F16 to F18. This dummy will not be added to the triple-difference regression, because the new dummy is full absorbed by the treatment zone–year fixed effects that characterize that regression. Because, as discussed in section 1, the quasi-experimental basis for the triple-difference estimate is shaky, I prefer the double-difference one augmented with this control for the satellite change.

I replicate and revise the HP19 nighttime light estimates following the template of the previous section. The initial revision step incorporates two changes already mentioned: fixing the coding error that causes some countries' post-treatment dummies to exceed 1; and working with the data at their original resolution of 1/120° instead of 1/10°, in order to assign treatment status more

precisely.[25,26]

The HP19 double-difference result proves robust to fixing the bug, increasing the resolution, and widening the treatment and control zones, as shown in the first column of Table 7. Indeed, the coefficients rise and the standard errors shrink. Similarly, the triple-difference estimate remains positive under the revisions, and, when fiber lines are the distance referent, highly significant statistically (column 3).

However, adding the treatment zone–post-2009 dummy to the double-difference regression—the specification preferred here to combat confounding—drives the result to 0 or below, without increasing standard errors (column 2 of the table).

In the right half of Table 7, switching to the intercalibrated light data changes these results to a surprising degree. The double-difference estimate falls while the triple-difference one rises (compare columns 1 and 4 and columns 3 and 6). However, the results from the preferred specification (double-difference with control for the satellite change) are more stable (columns 2 and 5). The estimates remain small or negative. The highest point estimate from this specification, 0.008, is a third of the HP19 estimate of 0.024.

On the evidence presented here, it is impossible to rule out that the satellite switch accounts for most or all of the seeming impact on night light reported from HP19's double-difference regression. As for the triple-difference specification, its aggressive control already removes confounding from the satellite switch, but in a way that is so flexible that the estimator is no longer clearly a quasi-experiment.

---

[25] Many authors have developed intercalibrated versions of the DMSP-OLS data (Li et al. 2013; Wu et al. 2013; Hsu et al. 2015; Stathakis 206; Li and Zhou 2017; Zhao et al. 2022). The minimally arbitrary choice is to stick to the Earth Observation Group (Hsu et al. 2015) since it produces the non-intercalibrated data.

[26] I also fit a Poisson model to the untransformed light values (Correia, Guimarães, and Zylkin 2020). However, these results are sensitive to the coding of the darkest pixels. While the documentation for the light data states that values run between 1 and 63, the darkest pixels are coded with 0, and 1 never occurs. The Poisson results are much more sensitive than the linear results to how this bottom-coding is done. Were it computationally feasible, a fixed-effect, bi-censored tobit model might be more appropriate for untransformed light values, since the sensors fail to detect signals below some threshold and saturate at the high end.

**Table 7. Impact of submarine cable arrival on nighttime light**

| | Raw light values | | | Intercalibrated light values | | |
|---|---|---|---|---|---|---|
| Original, 0.005°/0.1° from lines | 0.024 | | 0.033 | | | |
| | (0.009) | | (0.017) | | | |
| Observations | | 80,360 | | | | |
| + higher resolution, recalculated distance | 0.049 | –0.000 | 0.016 | 0.015 | 0.008 | 0.038 |
| | (0.003) | (0.003) | (0.004) | (0.006) | (0.003) | (0.003) |
| Observations | | 12,109,762 | | | 12,098,167 | |
| + Change to 0.1°/0.2° from lines | 0.040 | –0.003 | 0.011 | –0.003 | –0.003 | 0.033 |
| | (0.002) | (0.002) | (0.003) | (0.004) | (0.002) | (0.002) |
| Observations | | 20,685,168 | | | 20,666,997 | |
| + Change to 0.1°/0.2° from nodes | 0.069 | –0.012 | 0.004 | –0.009 | 0.007 | 0.050 |
| | (0.004) | (0.004) | (0.006) | (0.008) | (0.004) | (0.004) |
| Observations | | 10,537,478 | | | 10,527,659 | |
| Country-year FE | Yes | Yes | Yes | Yes | Yes | Yes |
| Treatment zone–grid cell FE | Yes | Yes | Yes | Yes | Yes | Yes |
| Treatment zone–post-2009 FE | No | Yes | No | No | Yes | No |
| Treatment zone–year FE | No | No | Yes | No | No | Yes |

Note: This table has the same format as Table 3. See notes to that table. Light values, ranging between 0 and 63, are linearly modeled after taking their arc-hyperbolic sines. Results in first row nearly match those in HP19 Table 9. Starting in the second row, the post-treatment indicator does not exceed 1 in countries that received cable multiple connections, correcting an error in the original. Raw and intertemporally calibrated values are from the Earth Observation Group. Standard errors, clustered by 0.1°×0.1° grid cell in the first two rows and 0.2°×0.2° in the last two, are in parenthesis.

**Figure 6. Evolution of association between stable nighttime light and proximity to fiber network**

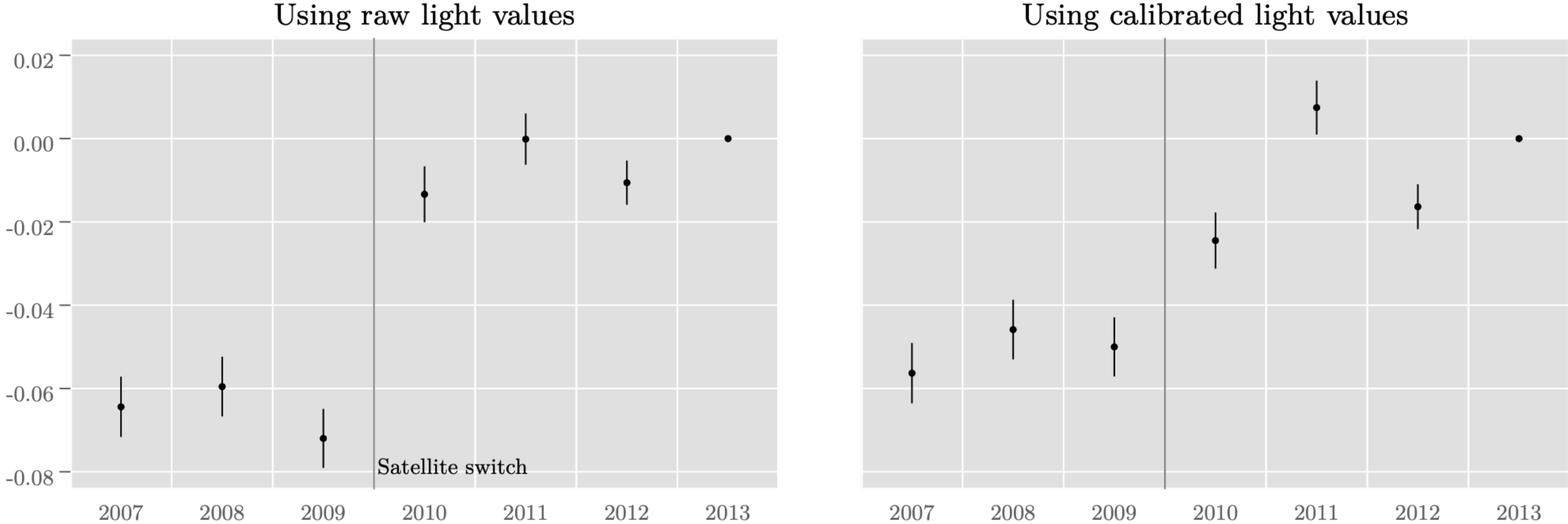

Notes: Each plot shows the estimated coefficients on products of the treated zone dummy and year dummy, along with 95% confidence intervals, in regressions whose specifications match those in the second row, third and sixth columns, of Table 7. The Defense Meteorological Satellite Program-Operational Linescan System stable light data set switches from satellite F16 to F18 between 2009 and 2010.

## 5 Conclusion

Reanalysis of HP19 weakens the evidence that undersea cable connections circa 2010 stimulated economic activity in Africa, in particular, increasing employment with a positive skill bias.

The largest reason for this revised view is the lack of robustness to changing the treatment and comparator zone radii from 0.005°/0.1° to 0.1°/0.2°. No definition of treatment and control zones can be conclusively preferred. But the new choices have several virtues. They are minimally arbitrary, as round numbers that define zones of equal width (0.1°). They have demonstrably higher fidelity to high-resolution indicators of broadband availability, and so should generate more statistical power. And they are more compatible with the resolution of the various geographic measurements used here. In other words, they step away from the pairing in HP19 of weaker identification and larger results. Under the new radii, the meta-analytically averaged estimates of individual-level employment impacts are smaller at 1.2 (0.9) points when fiber lines are the referent and 0.3 (0.6) when taking distance from nodes. The positivity in this bottom line derives mostly from regressions on the samples with the longest lags between pre- and post-treatment measurement—9 years, in the DHS data—thus the weakest claim to causal identification. Evidence of positive impact on firm-level employment disappears in the Ethiopia LMMIS surveys, as well as the WBES ones. Quasi-experimental evidence of impact on light emissions, as a proxy for economic activity, looks mostly like artifacts of a satellite changeover. Throughout, the signs of skill bias disappear.

Perhaps these results should not surprise. Economists have long rebutted the Luddite fallacy that technological advance destroys jobs. Generalizing, a reasonable prior is that any given technology's effects on aggregate employment are small and hard to detect.

## Appendices

# A Additional figures

**Figure A–1. Street map of part of Cotonou, Benin, with AfTerFibre representation of fiber line**

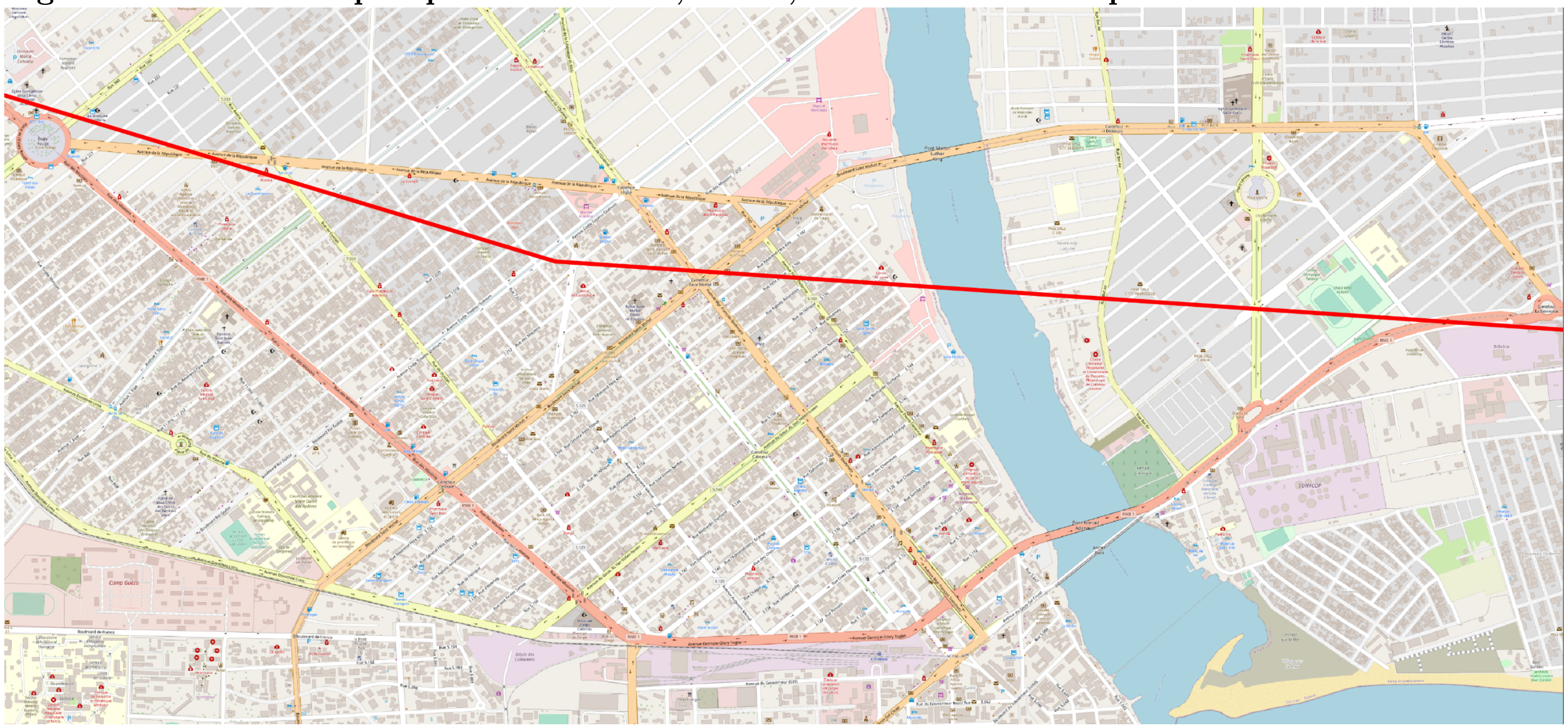

Notes: Map is from OpenStreetMaps (CC-BY-SA). Superimposed red line shows a fiber conduit as represented in the HP19 fiber network data. Most likely the actual fiber runs along the national highway that loops to the south, and only appears to deviate from the highway because its path is based on a digitization of the road route that, at this scale, appears imprecise. The maximum deviation is about 0.01°, twice the 0.005° used in HP19 to define the treatment zone along fiber lines.

**Figure A–2. Geocoding comparison, Akamai download cities**

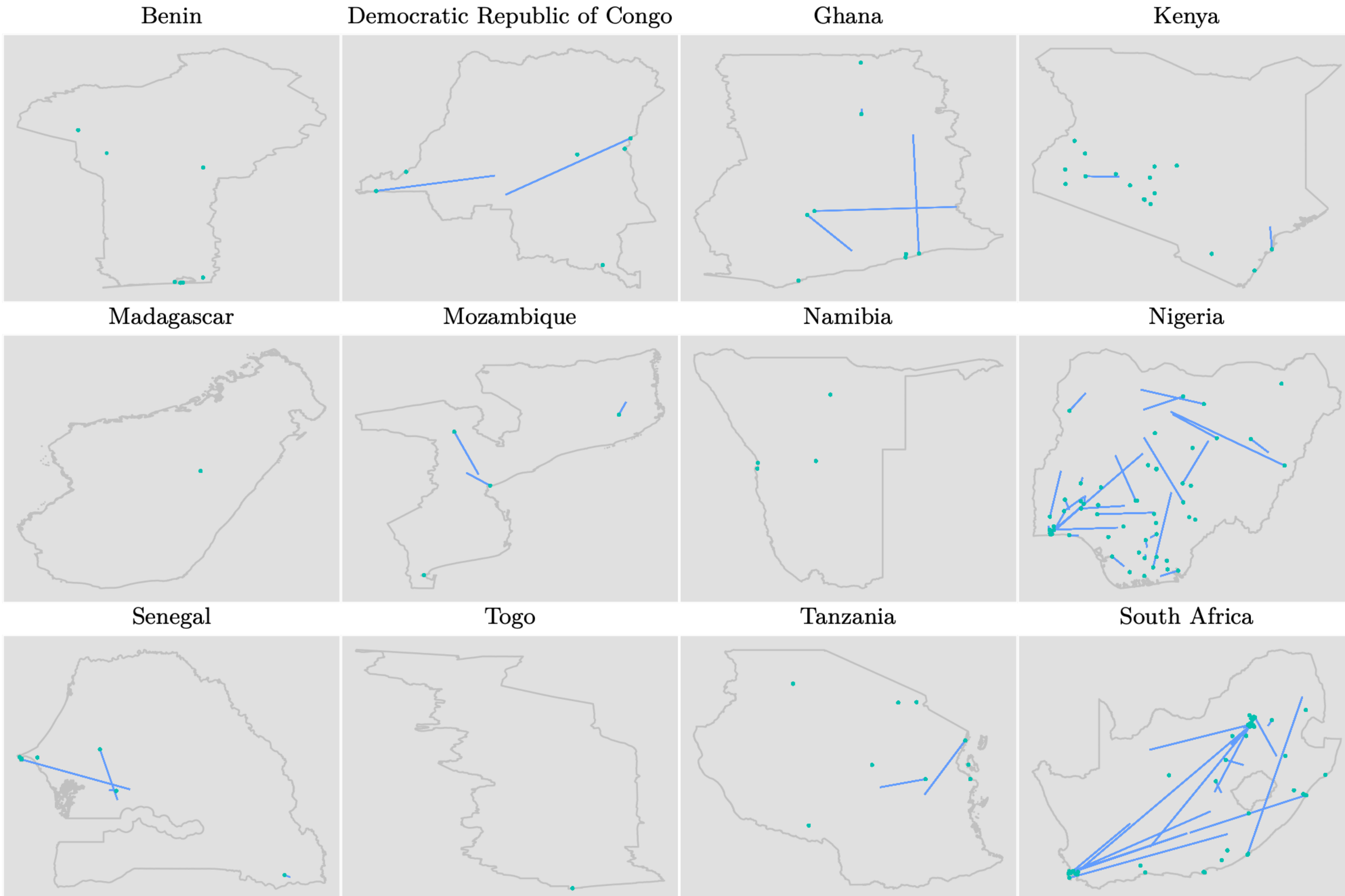

Notes: HP19 and this comment independently geocode the enumeration areas of the Akamai download locations. These plots compare the two codings by drawing blue lines between the points assigned to each enumeration area and marking the AidData ends with teal dots. Major discrepancies generate long lines with dots at the ends for the new geocodings, while close matches show just as dots. Scales are chosen separately for each axis and plot.

**Figure A–3. Geocoding comparison, Afrobarometer household survey locations**

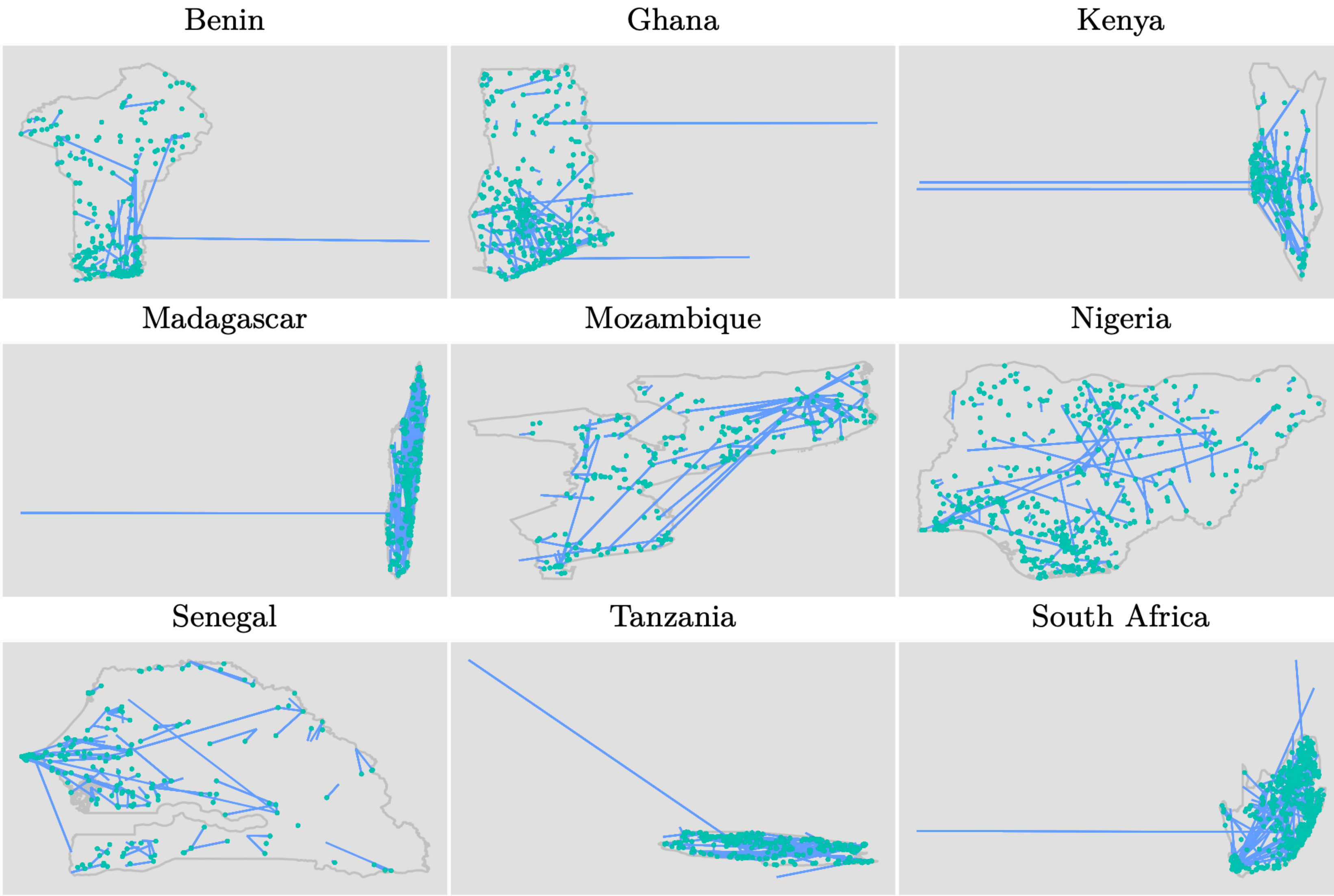

Notes: HP19 and AidData (BenYishay et al. 2017) independently geocode the enumeration areas of the Afrobarometer surveys. These plots compare the two codings; see notes to previous figure. Samples limited to locations with AidData precision code=1.

**Figure A–4. Treatment classification comparison, Akamai household survey locations**

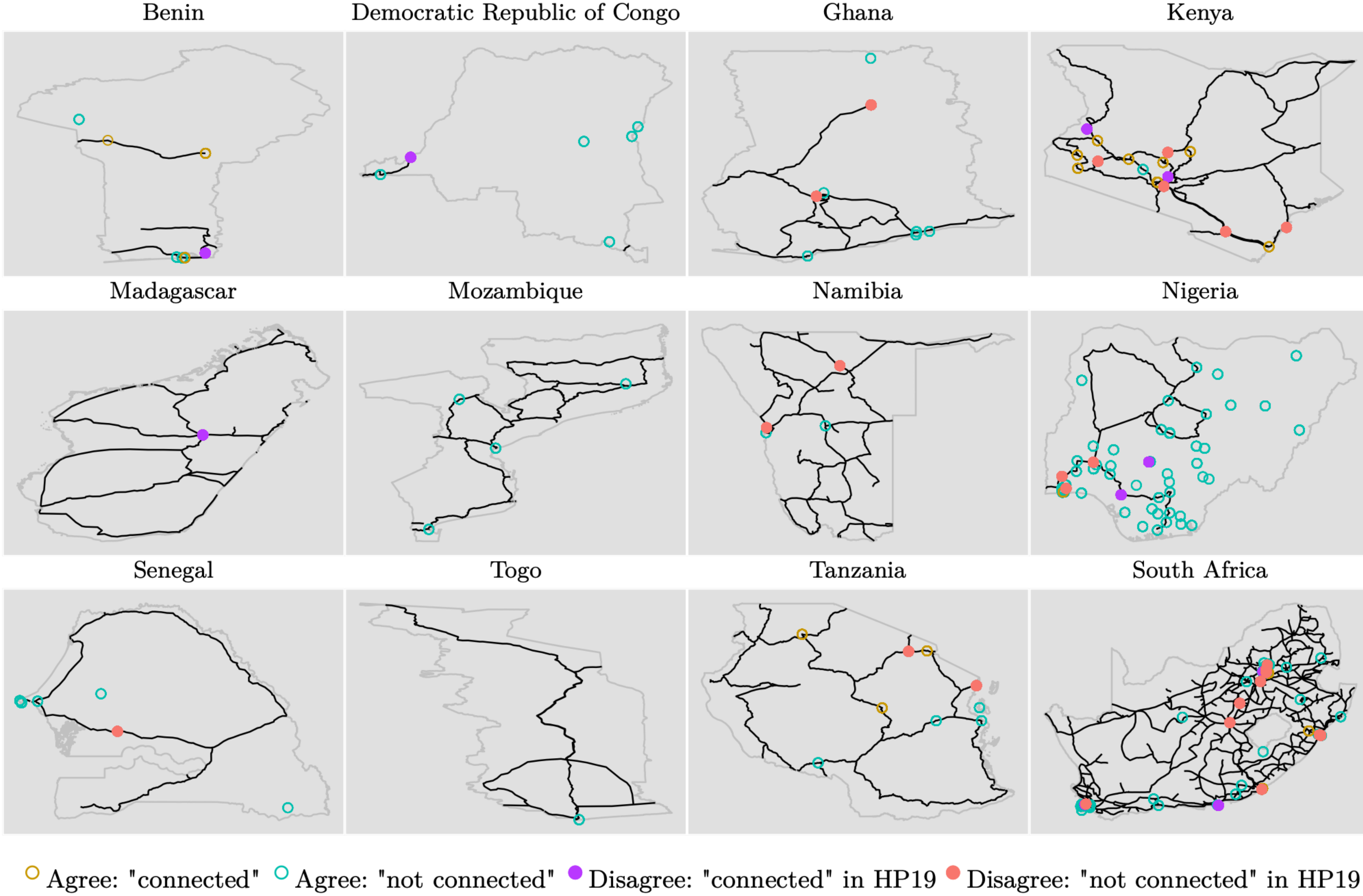

Notes: See notes to Figure 3.

**Figure A–5. Treatment classification comparison, DHS household survey locations**

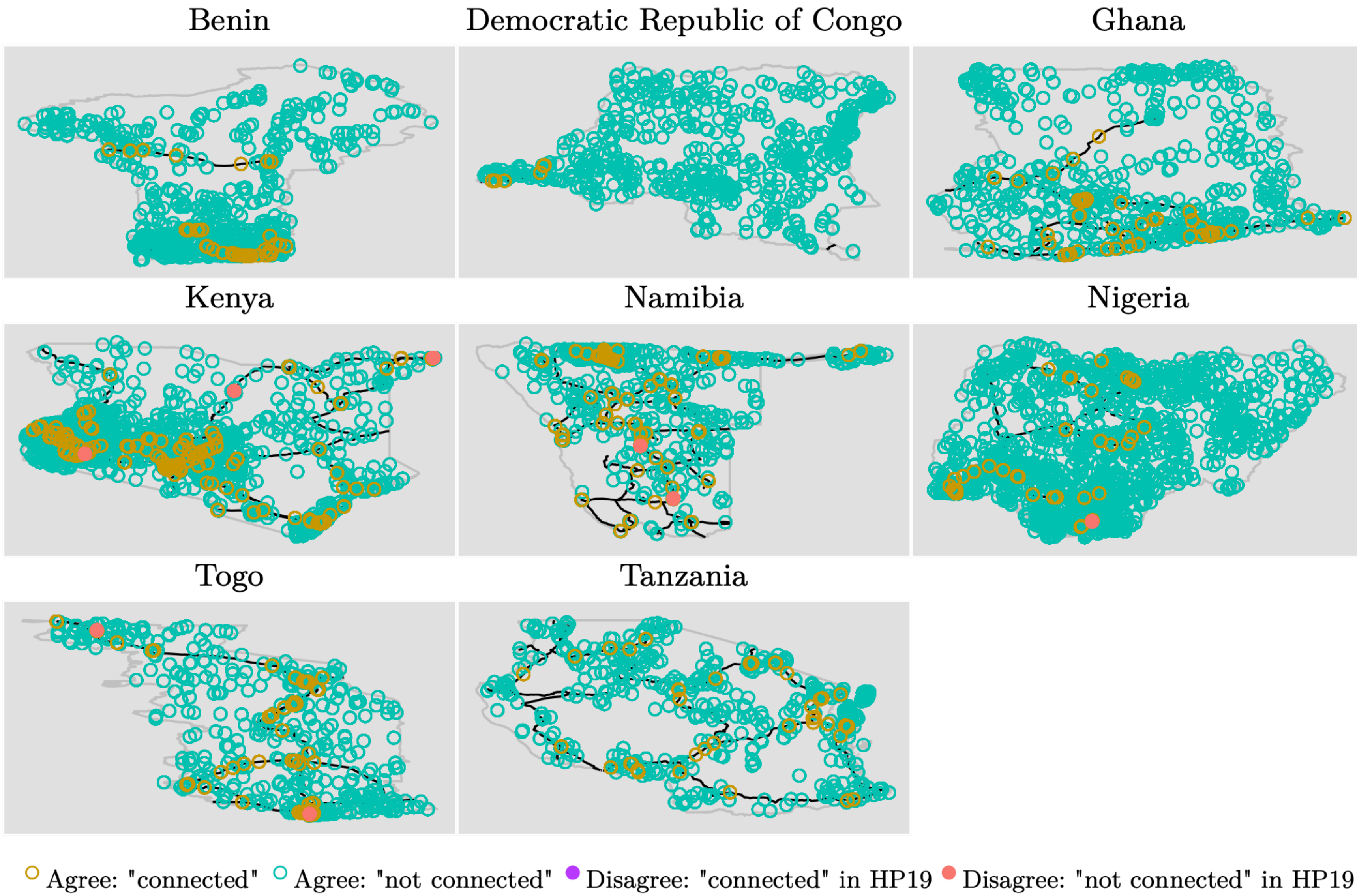

Notes: See notes to Figure 3. There are few disagreements because the HP19 and new classifications use location data provided by the DHS program.

**Figure A–6. Treatment classification comparison, South Africa QLFS enumeration areas**

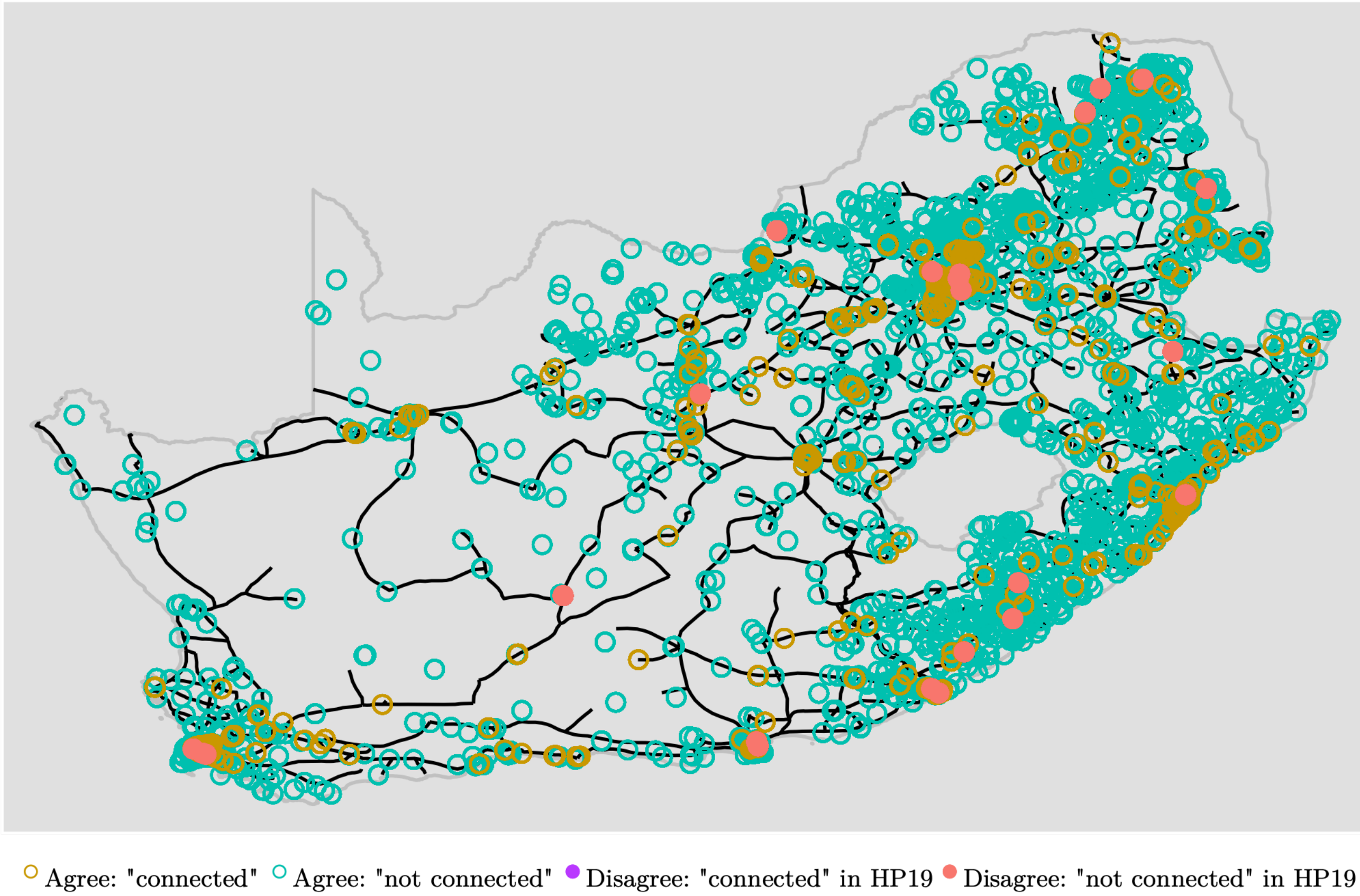

Notes: See notes to Figure 3.

**Figure A–7. Treatment classification comparison, Ethiopia LMMIS firm survey locations**

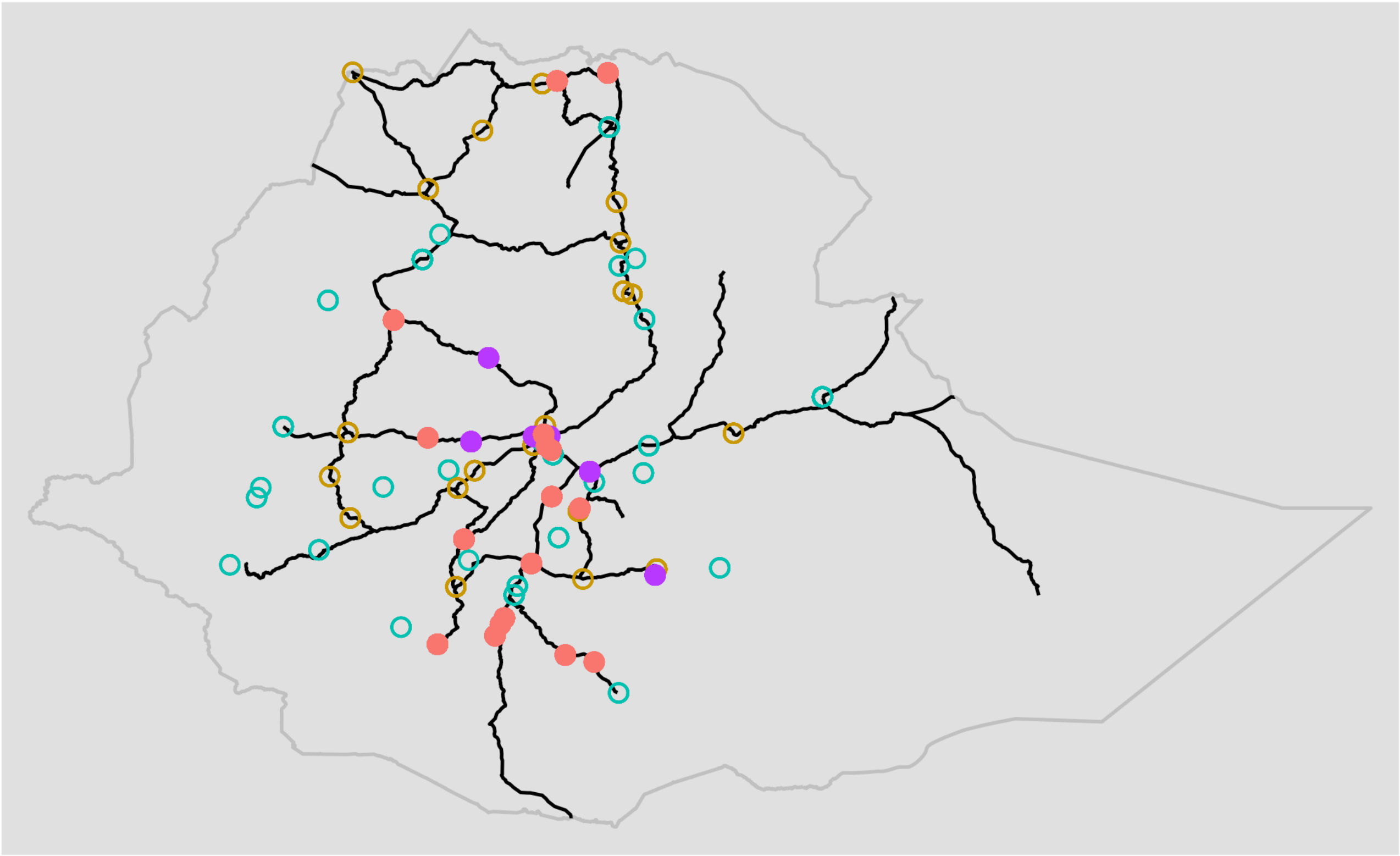

Notes: See notes to Figure 3. The 77 towns and woredas in the data are manually re-geocoded.

# B The econometrics of narrowing the treatment zone

For an exposition on the interaction of measurement error and endogeneity bias, we simplify from difference-in-differences to differences. For an observation $i$, conceive of the distance $d_i^*$ from the backbone as a signed quantity, or coordinate; for example, if a line runs north-south, points to the east could have positive distance and those to the west, negative. Distance is measured with error $\lambda_i^*$, yielding the observed distance

$$d_i = d_i^* + \lambda_i^*$$

To dichotomize the notion of treatment, given a treatment zone radius $r$, we define true and estimated treatment dummies,

$$t_i^* = \mathbf{1}\{|d_i^*| < r\}$$
$$t_i = \mathbf{1}\{|d_i| < r\}$$

Their difference,

$$\lambda_i = t_i - t_i^*$$

is the error in the measurement of the dichotomous treatment variable.[27]

Assume that in the $r \to 0$ limit,

$$\Pr[t_i^* = 1]\,, \Pr[t_i = 1]\,, \Pr[t_i = 1 | t_i^* = 1] \text{ are all } O(r) \tag{1}$$

For example, when $r$ is an extremely narrow 1 meter, the probability that an observation that is actually in the treatment zone will be measured as such ($\Pr[t_i = 1 | t_i^* = 1]$) will be close to zero; and if $r$ is halved, so will that conditional probability, approximately.

The outcome depends causally on distance according to

$$y_i = f(d_i^*) + \nu_i$$

where $f$ is a smooth, even function and $\mathrm{E}[\nu_i] = 0$. Given a value of $r$, such as 0.005°, the outcome equation for the associated *dichotomous* treatment indicator

[27] As $t_i$ and $t_i^*$ are dummies, the measurement error is non-classical, meaning that $\mathrm{E}[\lambda_i t_i^*] \neq 0$ (Aigner 1973). In particular, $\mathrm{E}[\lambda_i t_i^*] < 0$ because a 0 can only be mis-measured as 1 and vice versa.

is

$$y_i = \beta t_i^* + \epsilon_i \tag{2}$$

where $\epsilon_i$ is also a mean-zero error and

$$\beta = \mathrm{E}[f(d_i^*)|t_i^* = 1] - \mathrm{E}[f(d_i^*)|t_i^* = 0]$$

We assume that all variables are iid, which often allows us to drop $i$ subscripts without introducing ambiguity. But, crucially, we allow the estimated treatment dummy $t$ to be correlated with the true error $\epsilon$, whether because the true treatment dummy $t^*$ or the measurement error $\lambda_i$ is also correlated with $\epsilon$.

Let $\hat{\beta}_N$ be the result of ordinary least squares (OLS) regression of $y$ on $t$ in $N$ observations from this process. Substituting the above definitions into the formula for OLS and taking the $N \to \infty$ limit while fixing $r$ gives

$$\operatorname{plim} \hat{\beta}_n = \beta \frac{\mathrm{Cov}[t, t^*]}{\mathrm{Var}[t]} + \frac{\mathrm{Cov}[t, \epsilon]}{\mathrm{Var}[t]} \tag{3}$$

It follows from (1) that

$$\begin{aligned}
\mathrm{Cov}[t, t^*] &= \mathrm{E}[tt^*] - \mathrm{E}[t]\,\mathrm{E}[t^*] \\
&= \Pr[t^* = 1]\Pr[t = 1|t^* = 1] - \Pr[t^* = 1]\Pr[t = 1] \\
&= O(r^2) \\
\mathrm{Var}[t] &= \mathrm{E}[t^2] - \mathrm{E}[t]^2 \\
&= \Pr[t = 1] - \Pr[t = 1]^2 \\
&= O(r)
\end{aligned}$$

Thus the first term on the right of (3)—the exogenous identification component—goes to zero as $r \to 0$. But the second term, the endogenous component, in general does not. In particular, since the latter is the (infeasible) regression of $\epsilon$ on the dummy $t$, we have

$$\lim_{r \to 0} \frac{\mathrm{Cov}[t, \epsilon]}{\mathrm{Var}[t]} = \lim_{r \to 0} \{\mathrm{E}[\epsilon||d| < r] - \mathrm{E}[\epsilon||d \geq r]\} = \mathrm{E}[\epsilon|d = 0] - \mathrm{E}[\epsilon] = \mathrm{E}[\epsilon|d = 0]$$

If the measured distance $d$ contains endogenous measurement error, then by definition it contains information about $\epsilon$, and the last expression above is $O(1)$. So

$$\operatorname{plim} \hat{\beta}_n = \mathrm{E}[\epsilon|d = 0]$$

We find that narrowing the treatment zone below the resolution of most the data sources attenuates the exogenous component of $\hat{\beta}_n$—the first term on the right of (3)—toward zero, as in classical measurement error theory. But it does not attenuate the endogenous component, embodied in that second term. While the endogenous component may be second order when the treatment radius $r$ is large, it can come to dominate when $r$ is small. As a practical matter, the estimate $\hat{\beta}_n$ becomes less robust to endogeneity.